\RequirePackage{fix-cm}
\documentclass[smallextended]{svjour3}       
\smartqed

\usepackage{hyperref}
\usepackage{graphicx}
\usepackage{amsmath}
\usepackage{amssymb}
\usepackage{dsfont}
\usepackage[english]{babel}
\usepackage[latin1]{inputenc}
\usepackage[T1]{fontenc}

\newcommand{\bc}{\mathbf{c}}

\newcommand{\bT}{\mathbf{T}}

\newcommand{\bI}{\mathbf{I}}
\newcommand{\bmu}{\boldsymbol{\mu}}
\newcommand{\bsx}{\boldsymbol{x}}

\newcommand{\bsw}{\boldsymbol{w}}

\newcommand{\bst}{\boldsymbol{t}}

\newcommand{\bSigma}{\boldsymbol{\Sigma}}

\newcommand{\btheta}{\boldsymbol{\theta}}
\newcommand{\bsbeta}{\boldsymbol{\beta}}

\newcommand{\bal}{\boldsymbol{\alpha}}

\begin{document}

\title{Model-based clustering and segmentation of time series with changes in regime}


\author{Allou Samé         \and
        Faicel Chamroukhi \and Gérard Govaert \and Patrice Aknin
}


\institute{Allou Samé, Faicel Chamroukhi, Patrice Aknin \at
Institut Français des Sciences et Technologies des Transports,\\
de l'Aménagement et des Réseaux (IFSTTAR) \\
2 rue de la Butte Verte, 93166 Noisy-le-grand Cedex \\
\email{same@inrets.fr}
\and
Gérard Govaert \at
Université de Technologie de Compiègne (UTC)\\
UMR CNRS 6599, Heudiasyc\\
Centre de Recherches de Royallieu,
BP 20529, F-60205 Compiègne Cedex\\
\email{gerard.govaert@utc.fr}}
\date{Received: date / Accepted: date}

\maketitle

\begin{abstract}
Mixture model-based clustering, usually applied to multidimensional data, has become a popular
approach in many data analysis problems, both for its good statistical
properties and for the simplicity of implementation of the Expectation-Maximization (EM) algorithm. 
Within the context of a railway application, this paper introduces a novel mixture model for dealing with time series that are subject to changes in regime. The proposed approach consists in modeling each cluster by a regression model in which the polynomial coefficients vary according to a discrete hidden process. In particular, this approach makes use of logistic functions to model the (smooth or abrupt) transitions between regimes. The model parameters are
estimated by the maximum likelihood method solved by an
Expectation-Maximization algorithm. The proposed approach can also
be regarded as a clustering approach which operates by finding groups of
time series having common changes in regime. In addition to
providing a time series partition, it therefore provides a time series
segmentation. The problem of selecting the optimal numbers of clusters and segments is solved by means of the Bayesian Information Criterion (BIC). The proposed approach is shown to be efficient
using a variety of simulated time series and real-world time series of electrical power
consumption from rail switching operations.


\end{abstract}

\section{Introduction}
\label{intro}
The application which gave rise to this study is an application for diagnosing
problems in rail switches, that is to say the mechanisms which enable trains to change
tracks at junctions. One preliminary task
in the diagnostic process is identifying groups of switching
operations that have similar characteristics, and this is accomplished by
performing clustering on the time series of electrical power consumption,
acquired during various switching operations.
This kind of data is referred to
in other contexts as longitudinal data \cite{longitudinal}, signals, or curves \cite{gafney-regression-mixture}.

The approach adopted in this paper is mixture model-based clustering \cite{banfield,celeux}, which has successfully been applied in numerous domains \cite{maclachlan10}, and which provides, by means of the Expectation-Maximization algorithm \cite{demp77}, an efficient implementation framework. Typical extensions of mixture models for time series include regression mixture models \cite{gafney-regression-mixture} and random effect regression mixture models \cite{sugar,gafney-random-effect,ng2006,liu09}. These approaches are based on a projection of the original time series into a space with fewer dimensions, defined by polynomial or spline basis functions. Other approaches that combine Autoregressive Moving Average (ARMA) methods and the Expectation-Maximization algorithm have also been proposed \cite{arma_mix}.  Although these approaches can be seen as an efficient way of classifying time series, all of them use a constant dynamic within each cluster; in other words, the regressive or autoregressive coefficients of the clusters do not vary with time.

However, the time series studied in this paper are subject to various changes in regime (see figure \ref{real signals}) as a result of the successive mechanical movements that are involved in a
switching operation. Within this particular context, a specific regression model has been proposed in \cite{cham2} to deal with regime changes in time series. The model in question is a regression model in which the polynomial coefficients may vary according to a discrete hidden process, and which uses logistic functions to model the (smooth or abrupt) transitions between regimes. In this paper we extend this regression model
to a finite mixture model, where each cluster is represented by its own "hidden process regression model".

This paper is organized as follows. We first present a brief
review of the regression mixture model for time series clustering.
Then, we detail the proposed mixture model
and its parameters estimation via the Expectation-Maximization (EM)
algorithm \cite{demp77}. Section 5 illustrates the performances of the
proposed approach using simulated examples and real-world time series
from an application in the railway sector.

The time series to be classified takes the form of an independent
random sample $(\bsx_1,\ldots,\bsx_n)$ where each series $\bsx_i$
consists of a vector of $m$ random real values
$(x_{i1},\ldots,x_{im})$ observed over the fixed time grid
$\bst=(t_{1},\ldots,t_{m})$, with $t_1<t_2<\ldots<t_n$. The unobserved clusters corresponding to
$(\bsx_1,\ldots,\bsx_n)$ will be denoted as $(z_1,\ldots,z_n)$, where
$z_i\in \{1,\ldots,K\}$.

\section{Regression mixture model for time series clustering}
\label{sec:1}
This section briefly recalls the regression mixture model, as
formulated by Gafney and Smith \cite{gafney-regression-mixture}, in
the context of times series clustering.

\subsection{Definition of the regression mixture model}
Unlike standard vector-based mixture models, the density
of each component of the regression mixture is represented by a polynomial "mean
series" (or mean curve) parameterized by a vector of regression coefficients and a noise variance.

The regression mixture model therefore assumes that each series $\bsx_i$ is distributed
according to the conditional mixture density
\begin{equation}\label{melange 1}
f(\bsx_i|\bst;\btheta) = \sum_{k=1}^K \pi_k\,
\mathcal{N}(\bsx_i;\bT\bsbeta_k,\sigma^2_k\bI),
\end{equation}
where $\btheta=(\pi_1,\ldots,\pi_K,\bsbeta_1,\ldots,\bsbeta_K,\sigma^2_1,\ldots,\sigma^2_K)$
is the complete parameter vector, the $\pi_k$ are the proportions of the mixture satisfying $\sum_{k=1}^K\pi_k=1$, $\bsbeta_{k}$ and $\sigma^2_{k}$ are respectively the
the $(p+1)$-dimensional coefficient vector of the
$k$th regression model and the associated noise variance. The matrix $\bT=(T_{uj})$ is a $m\times (p+1)$ Vandermonde matrix verifying $T_{uj}=t^{u-1}_j$ for all ${1\leq j\leq m}$ and ${1\leq u\leq (p+1)}$, and
$\mathcal{N}(\cdot;\bmu,\bSigma)$ is the Gaussian density with mean vector
$\bmu$ and covariance matrix $\bSigma$.

\subsection{Fitting the model}
\label{sec:2}

Assuming that the sample $(\bsx_1,\ldots,\bsx_n)$ is independent,
the parameter vector $\btheta$ is estimated by maximizing the
conditional log-likelihood
\begin{eqnarray}
  \mathcal{L}(\btheta) = \log \sum_{i=1}^nf(\bsx_i|\bst;\btheta)
   = \sum_{i=1}^n \log \sum_{k=1}^K \pi_k \mathcal{N}(\bsx_i;\bT\bsbeta_k,\sigma^2_k\bI)
\end{eqnarray}
via the Expectation-Maximization (EM) algorithm initiated by
Dempster, Laird and Rubin \cite{demp77}.

Once the parameters have
been estimated, a time series partition  is obtained by assigning
each series $\bsx_i$ to the cluster having the highest posterior
probability

\begin{equation}
  p(z_i=k|\bst,\bsx_i;\btheta) = \frac{\pi_k \mathcal{N}(\bsx_i;\bT\bsbeta_k,\sigma^2_k\bI)}{\sum_{h=1}^K \pi_h\mathcal{N}(\bsx_i;\bT\bsbeta_h,\sigma^2_h\bI)}\cdot
\end{equation}

\section{Clustering time series with changes in regime}

\subsection{The global mixture model}

As with the standard regression mixture model, the mixture model introduced for clustering time series with changes in regime assumes that the series $\bsx_i$ are independently generated according to the global
mixture model
\begin{equation}\label{melange 2}
f(\bsx_i|\bst;\btheta) = \sum_{k=1}^K \pi_k
f_k(\bsx_i|\bst;\btheta_k),
\end{equation}
where $\btheta=(\pi_1,\ldots,\pi_K,\btheta_1,\ldots,\btheta_K)$, $\pi_1,\ldots,\pi_K$ denote the
proportions of the mixture, and $\btheta_k$ the parameters of
the different component densities $f_k$. The main difference between the model proposed here and Gafney and Smith's regression mixture model \cite{gafney-regression-mixture} lies in the definition of the component densities $f_k$, described in the following section.

\subsection{Definition of the mixture components}

We assume that the $k$th cluster, that is to say the time series corresponding to the component $f_k$ of the proposed mixture, is generated as follows. Given the cluster label $z_i=k$ and the fixed time vector $\bst$,
a time series $\bsx_i$ is  generated
according to a specific regression model which implicitly supposes
that there are $L$ $p$th order polynomial regression models involved
in the generation of $\bsx_i$. The assignment of the
$x_{ij}$'s to the different (sub) regression models is specified by a
hidden process denoted by $\bsw_i=(w_{i1},\ldots,w_{im})$, where
$w_{ij}\in\{1,\ldots,L \}$. Thus, given the cluster label $z_i=k$,
the individual observations $x_{ij}$ of a series $\bsx_i$ are
generated as follows:

\begin{equation}
\forall j=1,\ldots,m,\quad \left\{ \begin{array}{lll}
x_{ij}&=&\sum_{\ell=1}^L w_{ij\ell}\big(\bT'_j\bsbeta_{k\ell} +
\sigma_{k\ell}\varepsilon_{ij}\big)
\\\varepsilon_{ij} &\sim&
\mathcal{N}(0,1)\end{array}\right.,
\end{equation}
where $w_{ij\ell}=1$ if $w_{ij}=\ell$ and 0 otherwise. The parameters $\sigma_{k\ell}$ and $\bsbeta_{k\ell}$ are respectively the
noise standard deviation and  the $(p+1)$-dimensional coefficient vector of the
$\ell$th regression model of the $k$th cluster. $\bT'_{j}$ denotes the transpose of the vector $\bT_j=(1,t_{j},\ldots,t^p_{j})^T$.

The regression component labels $w_{ij}$ $(j=1,\ldots,m)$
are assumed to be generated according to the multinomial
distribution \linebreak
$\mathcal{M}(1,\pi_{k1}(t_{j};\bal_k),\ldots,\pi_{kL}(t_{j};\bal_k)),$
where
\begin{equation}
\pi_{k\ell}(t;\bal_k) =
\frac{\exp(\bal_{k{\ell}1}t+\bal_{k{\ell}0})}{\sum_{h=1}^{L}\exp(\bal_{kh1}t
+\bal_{kh0})}\cdot\label{logit_function}
\end{equation}
is a logistic function with
parameter vector $\bal_{k}=\{\bal_{k\ell};\ell=1,\ldots,L\}$ and\linebreak $\bal_{k\ell}=(\bal_{k{\ell}0},\bal_{k{\ell}1})$. A logistic function defined in this way ensures a smooth transition between the different polynomial regimes.
Thus, given $z_i=k$ and $t_j$, the individual
observations $x_{ij}$ of a series $\bsx_i$ are independently distributed according to
the mixture model given by
\begin{equation}\label{melange 3}
p(x_{ij}|t_{j};\btheta_k) = \sum_{\ell=1}^L
\pi_{k\ell}(t_{j};\bal_k)
\mathcal{N}(x_{ij};\bsbeta_{k\ell}^T\bT_{j},\sigma^2_{k\ell}).
\end{equation}
The density $f_k$ can thus be written as
\begin{equation}f_k(\bsx_i|\bst;\btheta_k)=\prod_{j=1}^{m} \sum_{\ell=1}^L
\pi_{k\ell}(t_{j};\bal_k)
\mathcal{N}(x_{ij};\bsbeta_{k\ell}^T\bT_{j},\sigma^2_{k\ell}).\end{equation}

\subsection{A cluster-segmentation model}
\begin{figure}[htbp]
\centering
\includegraphics[width=8.5cm, height=5.5cm]{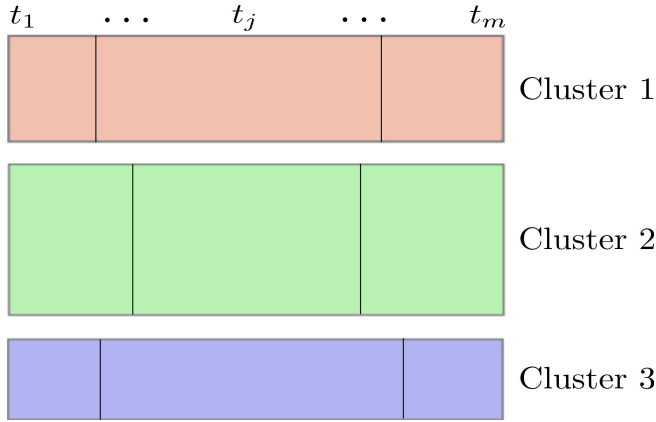}
\caption{Latent hierarchical structure of the proposed model with $K=3$ clusters: for each time series cluster, the vertical lines define a segmentation into $L=3$ segments}
\label{latent_struct}
\end{figure}

The proposed model leads to the segmentation $\textbf{E}_k=(E_{k\ell})_{\ell=1,\ldots,L}$ of the set of time series originating from the $k$th cluster, where
\begin{equation}
E_{k\ell} = \big \{t\in[t_1;t_m]\ /\   \pi_{k\ell}(t;\bal_k)=\max_{1\leq h \leq L}\pi_{kh}(t;\bal_k) \big\}.\label{segmentation}
\end{equation}
It can be proved that the set $E_{k\ell}$ is convex (see appendix \ref{appendix1}). Therefore, $\textbf{E}_k$ is a segmentation into contiguous parts of $\{t_1,\ldots t_m\}$. Figure \ref{latent_struct} illustrates the latent structure of the proposed model with $K=3$ and $L=3$.

\subsection{Parameter estimation via the EM algorithm}

The parameters of the proposed model are estimated by maximizing the
conditional log-likelihood defined by
\begin{eqnarray}
\mathcal{L}(\btheta) &=& \sum_{i=1}^n \log f(\bsx_i|\bst;\btheta) \nonumber\\
&=&\sum_{i=1}^n \log \sum_{k=1}^K \pi_k \Big(\prod_{j=1}^{m}
\sum_{\ell=1}^L \pi_{k\ell}(t_{j};\bal_k)
\mathcal{N}(x_{ij};\bsbeta_{k\ell}^T\bT_{j},\sigma^2_{k\ell})\Big).\label{vraisemblance
generale}
\end{eqnarray}

The Expectation Maximization (EM) algorithm \cite{demp77}
is used for the maximization of this log-likelihood, a problem which cannot be solved
analytically. Let us recall that the EM algorithm requires a complete data specification,
whose log-likelihood can be maximized more easily than the observed data log-likelihood.
Here, the "complete data" are obtained by adding to each series
$\bsx_i$ its cluster membership $z_i$ and its assignment
process $\bsw_i=(w_{ij})_{j=1,\ldots,m}$ to the different
sub-regression models. Using the binary coding of $z_i$ and
$\bsw_{ij}$,
$$z_{ik}=\left\{\begin{tabular}{ll}1 & \mbox{ if $z_i=k$}\\0& \mbox{ otherwise}\end{tabular}\right. \quad \mbox{ and }\quad w_{ij\ell}=\left\{\begin{tabular}{ll}1 & \mbox{ if $w_{ij}=\ell$}\\0& \mbox{ otherwise,}\end{tabular}\right.$$ the
complete data log-likelihood can be written as
\begin{eqnarray}
\mathcal{L}_c(\btheta)&=& \sum_{i=1}^n\log p(\bsx_i,z_i,\bsw_i|\bst;\btheta)=\sum_{i=1}^n
\sum_{k=1}^K z_{ik}\log \pi_k + \nonumber\\&&\sum_{i=1}^n \sum_{j=1}^{m}
\sum_{k=1}^K \sum_{\ell=1}^L z_{ik}w_{ij\ell}\log \Big(
\pi_{k\ell}(t_{j};\bal_k)\mathcal{N}(x_{ij};\bsbeta_{k\ell}^T\bT_{j},\sigma^2_{k\ell})\!\Big).
\end{eqnarray}

Given an initial value of the parameter vector $\btheta^{(0)}$, the
EM algorithm alternates the two following steps until
convergence.

\subsubsection*{E-Step (Expectation)}
This step consists in evaluating the expectation of the complete
data log-likelihood conditionally on the observed data and the
current parameter vector $\btheta^{(q)}$, $q$ denoting the current iteration:
\begin{eqnarray}
Q(\btheta,\btheta^{(q)}) &=& E \Big[
\mathcal{L}_c(\theta)\big|\bst,\bsx_1,\ldots,\bsx_n;\btheta^{(q)}
\Big]=\sum_{i=1}^n \sum_{k=1}^K r^{(q)}_{ik}\log \pi_k +\nonumber\\&&
\sum_{i=1}^n \sum_{j=1}^{m} \sum_{k=1}^K \sum_{\ell=1}^L
\lambda^{(q)}_{ijk\ell}\log \Big(
\pi_{k\ell}(t_{j};\bal_k)\mathcal{N}(x_{ij};\bsbeta_{k\ell}^T\bT_{j},\sigma^2_{k\ell})\!\Big)
\end{eqnarray}
where
\begin{eqnarray}r^{(q)}_{ik}=E[z_{ik}|\bst,\bsx_i;\btheta^{(q)}] =\frac{\pi^{(q)}_k f_k(\bsx_i|\bst;\btheta^{(q)}_k)}{\sum_{h=1}^K\pi^{(q)}_{h}
f_{h}(\bsx_i|\bst;\btheta^{(q)}_{h})}\end{eqnarray}
is the posterior probability that time series $\bsx_i$
originates from cluster $k$, and
\begin{eqnarray}
\lambda^{(q)}_{ijk\ell} &=&
E[z_{ik}\,w_{ij\ell}|\bst,\bsx_i;\btheta^{(q)}]\nonumber\\
&=&\small\frac{\pi^{(q)}_k f_k(\bsx_i|\bst;\btheta^{(q)}_k)}{\sum_{h=1}^K\pi^{(q)}_{h}
f_{h}(\bsx_i|\bst;\btheta^{(q)}_{h})}\times\frac{\pi_{k\ell}(t_{j};\bal^{(q)}_k)\mathcal{N}(x_{ij};{\bsbeta^{(q)^T}_{k\ell}}\bT_{j},{\sigma^{2^{(q)}}_{k\ell}})}{\sum_{h=1}^L\pi_{kh}(t_{j};\bal^{(q)}_k)\mathcal{N}(x_{ij};{\bsbeta^{(q)^T}_{kh}}\bT_{j},{\sigma^{2^{(q)}}_{kh}})
}
\end{eqnarray}
is the posterior probability that $(t_{j},x_{ij})$
originates from the $\ell$th sub-regression model of cluster $k$.

\subsubsection*{M-Step (Maximization)}

This step consists in computing the parameter vector
$\btheta^{(q+1)}$ that maximizes the quantity
$Q(\btheta,\btheta^{(q)})$ with respect to $\btheta$. For our purposes this quantity can
be written as
$$Q(\btheta,\btheta^{(q)})=Q_1((\pi_k))+  Q_2((\bal_k))+ Q_3((\beta_{k\ell},\sigma^2_{k\ell})),$$
 where
\begin{eqnarray}
Q_1((\pi_k)) & =& \sum_{i=1}^n \sum_{k=1}^K r^{(q)}_{ik}\log \pi_k,\\
Q_2((\bal_k))&=&\sum_{i=1}^n \sum_{j=1}^{m} \sum_{k=1}^K
\sum_{\ell=1}^L \lambda^{(q)}_{ijk\ell}\log \Big(
\pi_{k\ell}(t_{j};\bal_k)\!\Big),\label{Q2}\\
Q_3((\beta_{k\ell},\sigma^2_{k\ell}))&=&\sum_{i=1}^n \sum_{j=1}^{m}
\sum_{k=1}^K \sum_{\ell=1}^L \lambda^{(q)}_{ijk\ell}\log
\Big(\mathcal{N}(x_{ij};\bsbeta_{k\ell}^T\bT_{j},\sigma^2_{k\ell})\!\Big).
\end{eqnarray}
$Q$ can thus be maximized by separately maximizing
the quantities $Q_1$, $Q_2$ and $Q_3$. As in the classical Gaussian mixture model, it can easily be
shown that the proportions $\pi_k$ that maximize $Q_1$ under the
constraint $\sum_{k=1}^K \pi_k=1$ are given by
\begin{equation}
\pi^{(q+1)}_k=\frac{\sum_{i=1}^n r^{(q)}_{ik}}{n}\cdot
\end{equation}

$Q_2$ can be maximized with respect to the $\bal_k$ by separately
solving $K$ weighted logistic regression problems:
\begin{equation} \bal^{(q+1)}_k=\arg \max_{\bal_k}\sum_{i=1}^n \sum_{j=1}^{m} \sum_{\ell=1}^L
\lambda^{(q)}_{ijk\ell}\log \Big(
\pi_{k\ell}(t_{j};\bal_k)\!\Big)\end{equation} through the well known
Iteratively Reweighted Least Squares (IRLS) algorithm \cite{irls,cham2}. Let us recall that the IRLS algorithm, which is generally used to estimate
the parameters of a logistic regression model, is equivalent to the following
Newton Raphson algorithm \cite{irls,cham2}:

\begin{equation}
\bal^{(v+1)}_k = \bal^{(v)}_k - \Big[\frac{\partial^2 Q_{2k}}{\partial \bal_k \partial \bal^T_k}\Big]^{-1}_{{\bal_k=\bal^{(v)}_k}} \Big[\frac{\partial Q_{2k}}{\partial \bal_k}\Big]_{{\bal_k=\bal^{(v)}_k}},
\end{equation}
where $$Q_{2k}= \sum_{i=1}^n \sum_{j=1}^{m}
\sum_{\ell=1}^L \lambda^{(q)}_{ijk\ell}\log
\pi_{k\ell}(t_{j};\bal_k).$$

Maximizing $Q_3$ with respect to $\bsbeta_{k\ell}$ consists
in analytically solving $K\times L$ weighted least-squares problems.
It can be shown that
\begin{equation}
\bsbeta^{(q+1)}_{k\ell} = \Big[ \bT' \big(\sum_{i=1}^n \Lambda^{(q)}_{i
k\ell}\big)\bT \Big]^{-1}\Big[  \bT \big(\sum_{i=1}^n
\Lambda^{(q)}_{i k\ell}\big)\bsx_i\Big],
\end{equation}
where $\Lambda^{(q)}_{i k\ell}$ is the $m\times m$ diagonal matrix
whose diagonal elements are \linebreak $\big\{\lambda^{(q)}_{ijk\ell} \ ;\
j=1,\ldots,m\big\}$. The maximization of $Q_3$ with respect to $\sigma^2_{k\ell}$ gives
\begin{equation}
\left(\sigma^2_{k\ell}\right)^{(q+1)}=\frac{\sum_{i=1}^n \big \|
\sqrt{\Lambda^{(q)}_{ik\ell}}\big(\bsx_i-\bT\bsbeta^{(q+1)}_{k\ell}\big)
\big \|^2}{\sum_{i=1}^n\mbox{trace}(\Lambda^{(q)}_{ik\ell}) },
\end{equation}
where  $\sqrt{\Lambda^{(q)}_{ik\ell}}$ is the $m\times m$ diagonal
matrix whose diagonal elements are \linebreak
$\big\{\sqrt{\lambda^{(q)}_{ijk\ell}} \ ;\ j=1,\ldots,m\big\}$ and $\| \cdot\|$ is the norm corresponding to the euclidian distance.

\subsubsection*{M-step for three parsimonious models}

\noindent{\bf Common segmentation for all clusters}

\noindent
In certain situations, the segmentation defined by the $\bal_k$ $(k=1,\ldots,K)$ may be constrained to be common for each cluster, that is $\bal_k=\bal$ $\forall k$. In that case, the quantity $Q_2$
to be maximized can be rewritten as:
\begin{equation}
Q_{2}(\bal)=\sum_{i=1}^n \sum_{j=1}^{m} \sum_{\ell=1}^L
\lambda^{(q)}_{ij\cdot\ell}\log \Big( \pi_{\ell}(t_{j};\bal)\!\Big),
\end{equation}
where $
\lambda^{(q)}_{ij\cdot\ell}=\sum^K_{k=1}\lambda^{(q)}_{ijk\ell}$. The IRLS algorithm can therefore be used to compute the parameter $\bal^{(q+1)}$, in the same way as for the unconstrained situation.

\pagebreak
\noindent{\bf Common variance for regression models from the same cluster}

\noindent
In other situations, it may be useful to constrain the regression models variances to be common within a same cluster. In that case,
$\sigma^2_{k\ell}=\sigma^2_k \ \forall k,\ell$. The
updating formula for the variance can thus be written as:
\begin{equation}
\left(\sigma^2_k
\right)^{(q+1)}=\frac{\sum_{i=1}^n\sum_{\ell=1}^L \left
\|
\sqrt{\Lambda^{(q)}_{ik\ell}}\left(\bsx_i-\bT\bsbeta^{(q+1)}_{k\ell}\right)
\right \|^2}{\sum_{i=1}^n\sum_{\ell=1}^L\mbox{trace}(\Lambda^{(q)}_{ik\ell})}\cdot
\end{equation}

\medskip
\noindent{\bf Common variance for all regression models}

\noindent
If the model variances are constrained to be common all regression models, we have
$\sigma^2_{k\ell}=\sigma^2 \ \forall k,\ell$. The
updating formula for the variance takes the form:
\begin{equation}
\left(\sigma^2
\right)^{(q+1)}=\frac{\sum_{i=1}^n\sum_{k=1}^K\sum_{\ell=1}^L \left
\|
\sqrt{\Lambda^{(q)}_{ik\ell}}\left(\bsx_i-\bT\bsbeta^{(q+1)}_{k\ell}\right)
\right \|^2}{n\times m}\cdot
\end{equation}

\subsection{Time series clustering, approximation and segmentation}
From the parameters estimated by the EM algorithm, a partition of the
time series can  easily be deduced by applying the maximum a posteriori
(MAP) rule \begin{equation}z_i=\arg \max_k r_{ik}.\end{equation}

The clusters "mean series" can be
approximated by the series $\bc_k=(c_{kj})$, with \begin{equation}c_{kj}=E[x_{ij}|t_j,z_i=k;\btheta]=\sum_{\ell=1}^L
\pi_{k\ell}(t_j;\bal_k)\bT'_j\bsbeta_{k\ell}.\end{equation}

Moreover, a segmentation $\textbf{E}_k=(E_{k\ell})_{\ell=1,\ldots,L}$ of the time series originating from the $k$th cluster can be derived from the estimated parameters by computing $E_{k\ell}$ as defined in equation \ref{segmentation}.

\subsection{Assessing the number of clusters, segments and the regression order}

In the context of mixture models and the EM algorithm, the natural
criterion for model selection is the Bayesian Information Criterion
(BIC) \cite{BIC}. Unlike for classical mixture regression models, three parameters
need to be tuned: the number of clusters $K$, the
number of segments $L$ and the degree $p$ of the polynomials. The BIC
criterion, in this case, can be defined by:
\begin{equation}
BIC(K,L,p)=L(\widehat{\btheta})-\frac{\nu(K,L,p)}{2}\log(n),
\end{equation}
where $\widehat{\btheta}$ is the parameter vector estimated by the
EM algorithm, and $\nu(K,L,p)$ is the number of free parameters of
the model. In the proposed model, the number of free parameters
\begin{equation}
\nu(K,L,p)=(K-1) + 2\,K(L-1) + L\, K(p+1) + L\, K
\end{equation} is the sum of the mixture proportions, the logistic functions parameters, the polynomial coefficients and the variances.

From a practical point of view, the maximum numbers $K_{max}$, $L_{max}$ and $p_{max}$ are first specified. Then, the EM algorithm is run for $K \in\{1,\ldots,K_{max}\}$, $L \in\{ 1,\ldots,L_{max}\}$ and $p\in\{1,\ldots,p_{max}\}$, and the BIC criterion is computed. The set $(K,L,p)$ with the highest value of BIC is taken to be right solution.

\section{Experimental study}

This section is devoted to an evaluation of the clustering accuracy of the proposed algorithm, carried out
using simulated time series and real-world time series from a
railway application. Results obtained from the proposed algorithm are compared with those provided by the clustering approach, based on the regression mixture described in section 2.
To measure the clustering accuracy, two criteria were used: the misclassification percentage between the true partition and the estimated partition, and the intra-cluster inertia
$
\sum_{k=1}^K\sum_{i=1}^n \widehat{z}_{ik}||\bsx_i-\widehat{\bc}_k||^2,
$
where $(\widehat{z}_{ik})$ and $\widehat{\bc}_k= (\widehat{c}_{kj})_{j=1,\ldots,m}$ represent respectively the binary partition matrix and the $k$th mean series estimated by each of the two compared algorithms:
\begin{itemize}
\item $\widehat{c}_{kj}=\sum_{\ell=1}^L \pi_{k\ell}(t_j;\bal_k) \bT'_j \bsbeta_{k\ell}$ for the proposed algorithm,
\item $\widehat{c}_{kj}= \bT'_j \bsbeta_k$ for the regression mixture EM algorithm.
\end{itemize}

\subsection{Experiments using simulated data}

\subsubsection{Simulation protocol and algorithms tuning}

The time series are simulated as follows: $n$ series of length $m$
are generated according to a mixture of $K$ clusters whose mean series
can be either polynomial or the sum of polynomials weighted by logistic functions.

The polynomial coefficients and variances are initialized as follows: $K$ series are randomly selected and segmented into $L$ regularly spaced segments; the
polynomial regression parameters are derived from a
$p$th order regression on each segment. The logistic regression
parameters are initialized to the null vector.
The initial polynomial
coefficients and variances of the regression mixture approach are obtained by performing a $p$th-order
regression on $K$ randomly drawn series. The proportions of the initial clusters are set to $1/K$
for all algorithms. Each algorithm starts with
20 different initializations and  the solution with the highest log-likelihood is selected.

\subsubsection{Comparison between the proposed model and the standard regression mixture}
\begin{figure}[htbp]
\centering
\begin{tabular}{cc}
 \multicolumn{2}{c}{\includegraphics[width=5cm,height=3cm]{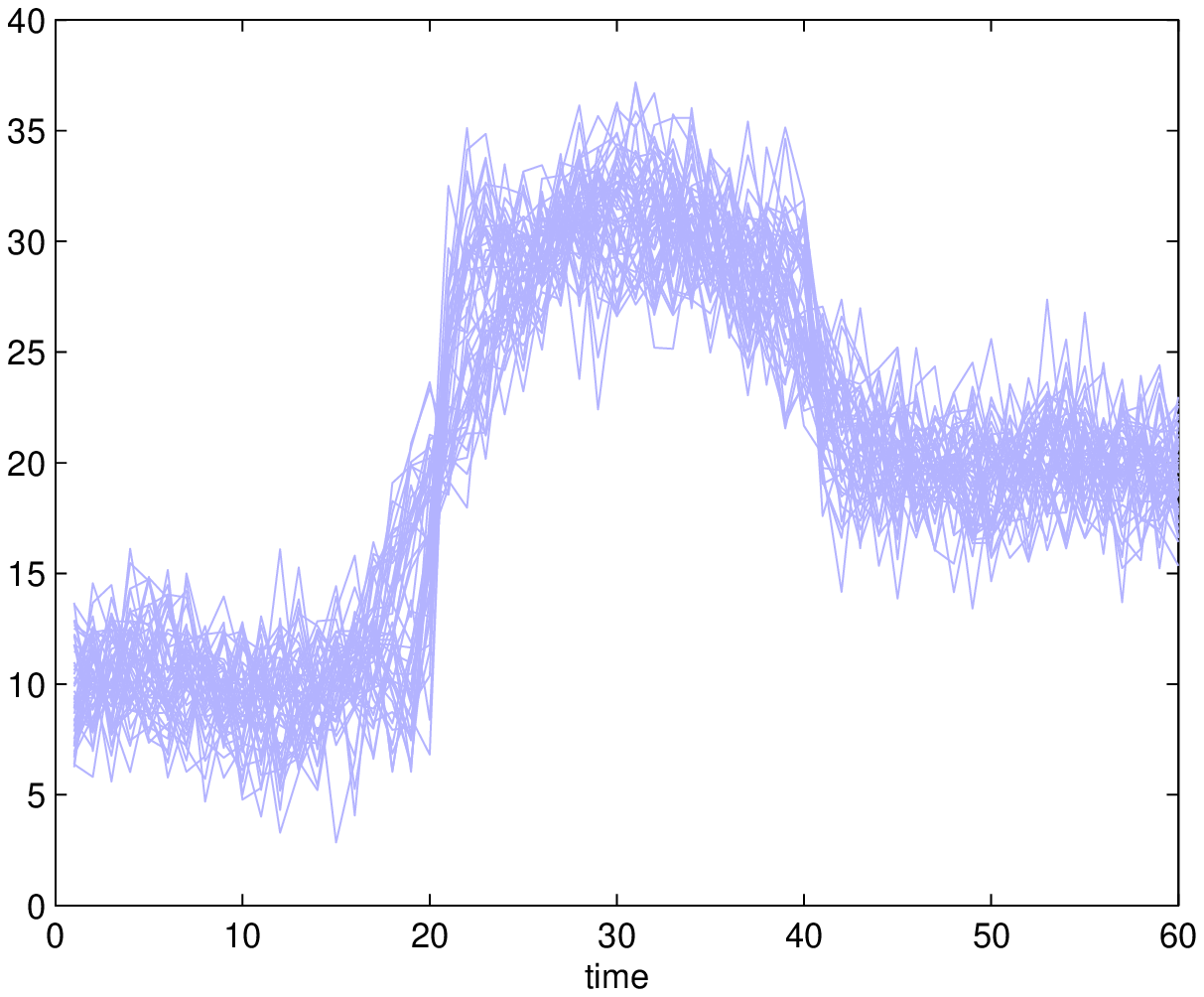}}\\\multicolumn{2}{c}{(a)}\\
 \includegraphics[width=5cm,height=3cm]{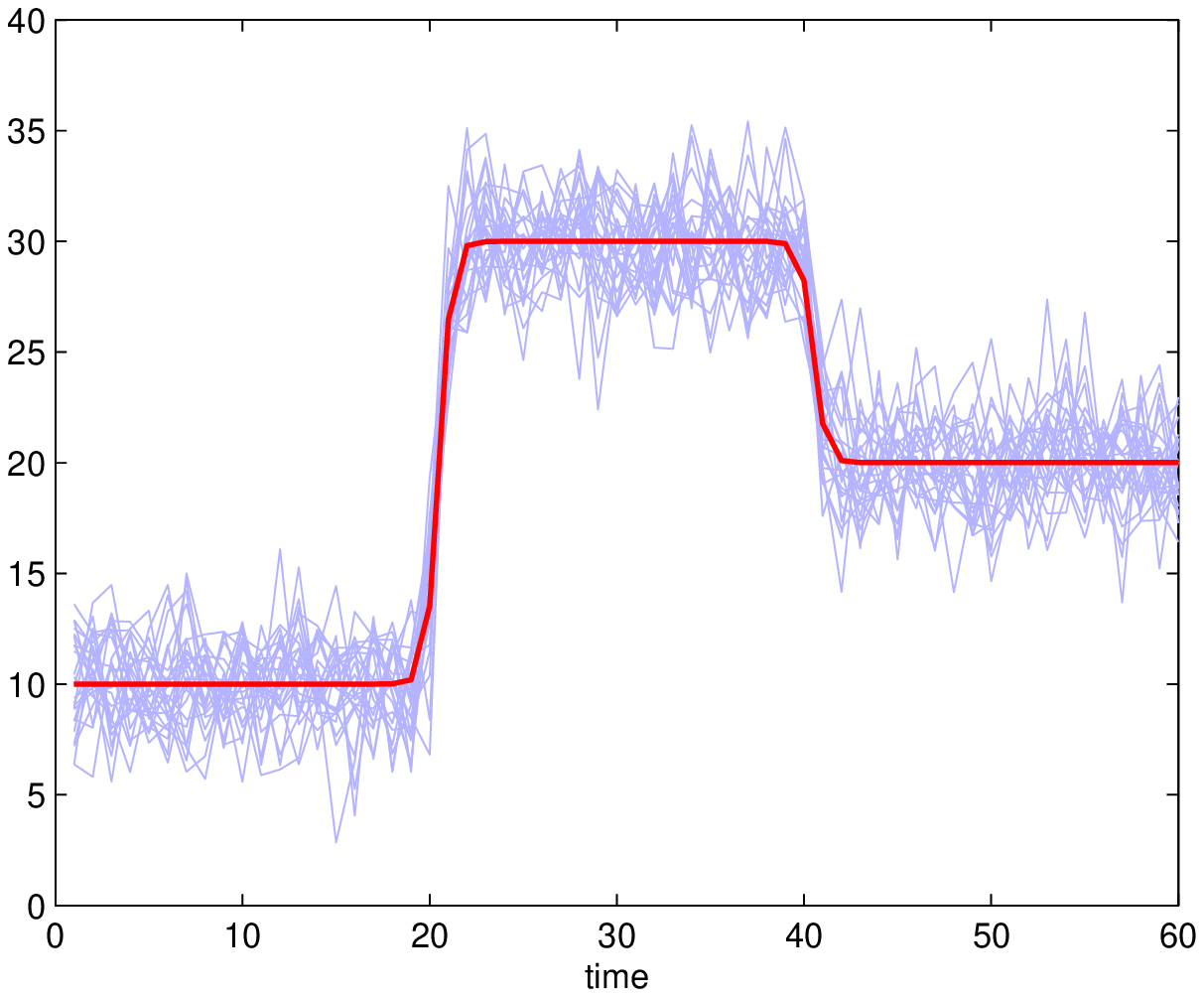}&
 \includegraphics[width=5cm,height=3cm]{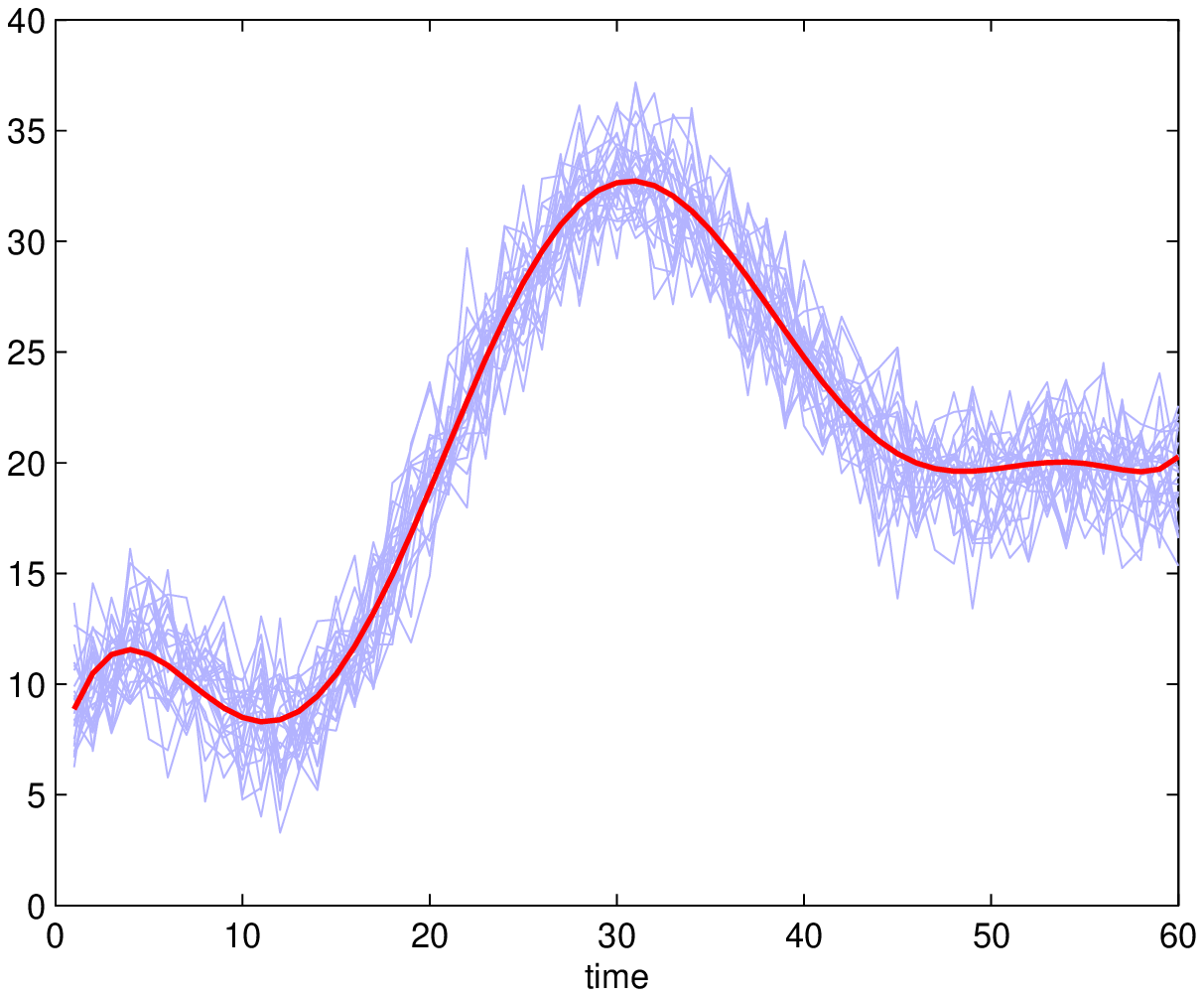}\\(b)&(c)\\
\end{tabular}
\caption{Example of $n=50$ simulated time series (a) and series corresponding to the two clusters, with their mean (b and c)}\label{simulation1}
\end{figure}

The experiments were performed in order to compare the relative performances of the proposed EM algorithm and the EM algorithm for standard regression mixtures. So as not to favor either method unduly, the data were generated without
reference either to the proposed model or to the regression mixture. Each data set, consisting of $n=50$ time series of length $m=60$, was simulated according to a mixture of $K=2$ clusters with equal proportions ($\pi_1=\pi_2=1/2$). The first cluster mean curve was built from three polynomials of degree $p=0$ weighted by logistic functions, while the second was a single polynomial of degree $p=8$. Values of the variance $\sigma^2_{k\ell}$ were chosen equal for each of the simulated sets of time series. The parameters of the mean curves are given in table \ref{tab1_param}, and figure \ref{simulation1} provides an illustration of time series simulated according to this model.

\begin{table}[htbp]
\centering
\caption{Clusters' mean series with their parameters}\label{tab1_param}
{\scriptsize
\begin{tabular}{lll}
\hline
Cluster&Mean series & \multicolumn{1}{l}{Parameters}\\\hline
$k=1$&$c_{1j}=\sum_{\ell=1}^3\pi_{1\ell}(t_j;\bal_1)\bT'_j\bsbeta_{1\ell} $&$\bsbeta_{11}=10\quad$ $\bal_{11}=(1039, -34.4)'$ \\
&&$\bsbeta_{12}=20\quad$ $\bal_{12}=(677, -16.7)'$ \\
&&$\bsbeta_{13}=30\quad$ $\bal_{13}=(0,0)'$ \\
\hline
$k=2$&$c_{2j}=\bT'_j\bsbeta_{2} $&$\bsbeta_{2}=(7.4,\,  1.9,\, -0.3,\, -2\times 10^{-3},\,2\times 10^{-4}, $\\
&&$-1.3\!\times\!10^{-4}, 3.2\!\times\!10^{-6}, -3.7\!\times\!10^{-8}, 1.6\!\times\!10^{-10})'$  \\
\hline
\end{tabular}}
\end{table}

\begin{figure}[htbp]
\centering
\begin{tabular}{ccc}
\includegraphics[width=6cm]{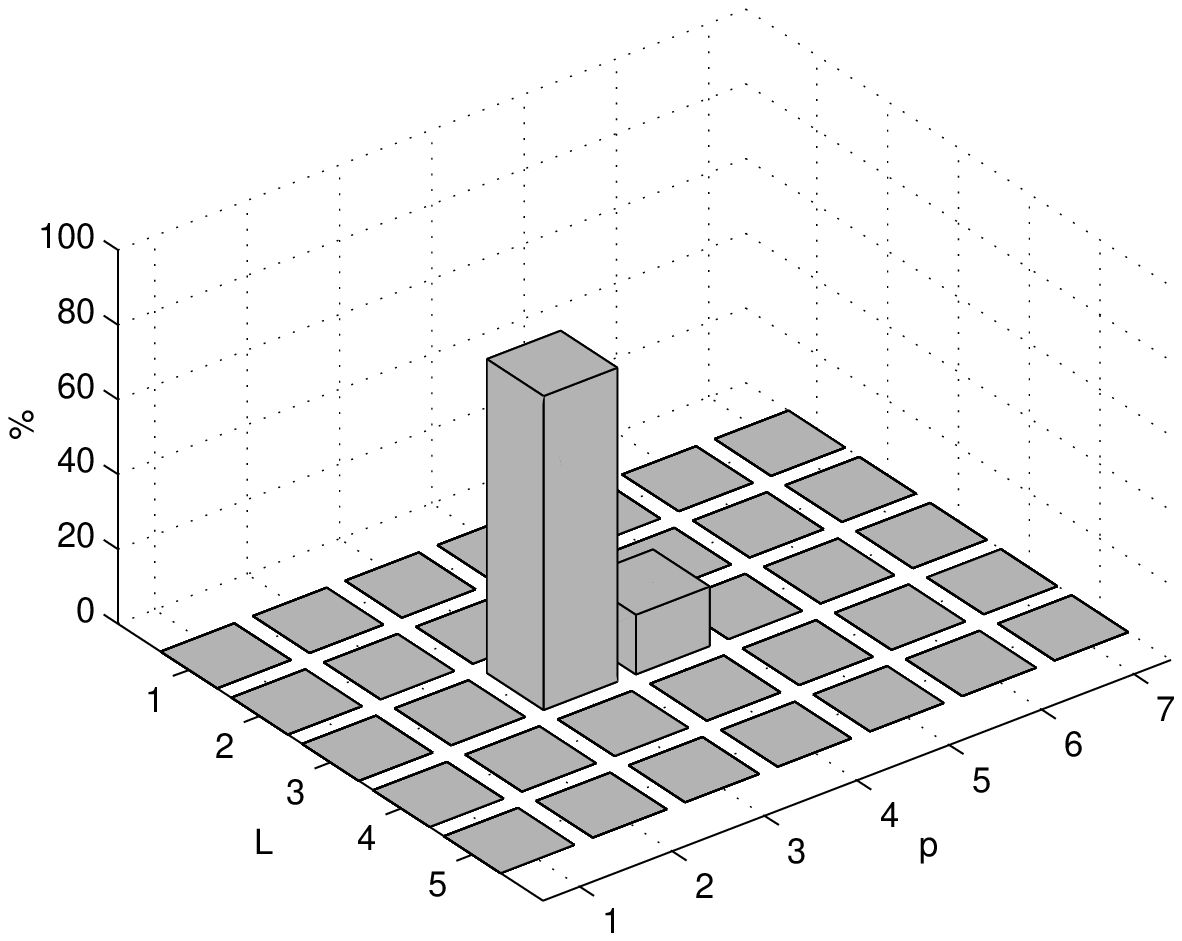} & \includegraphics[width=5cm]{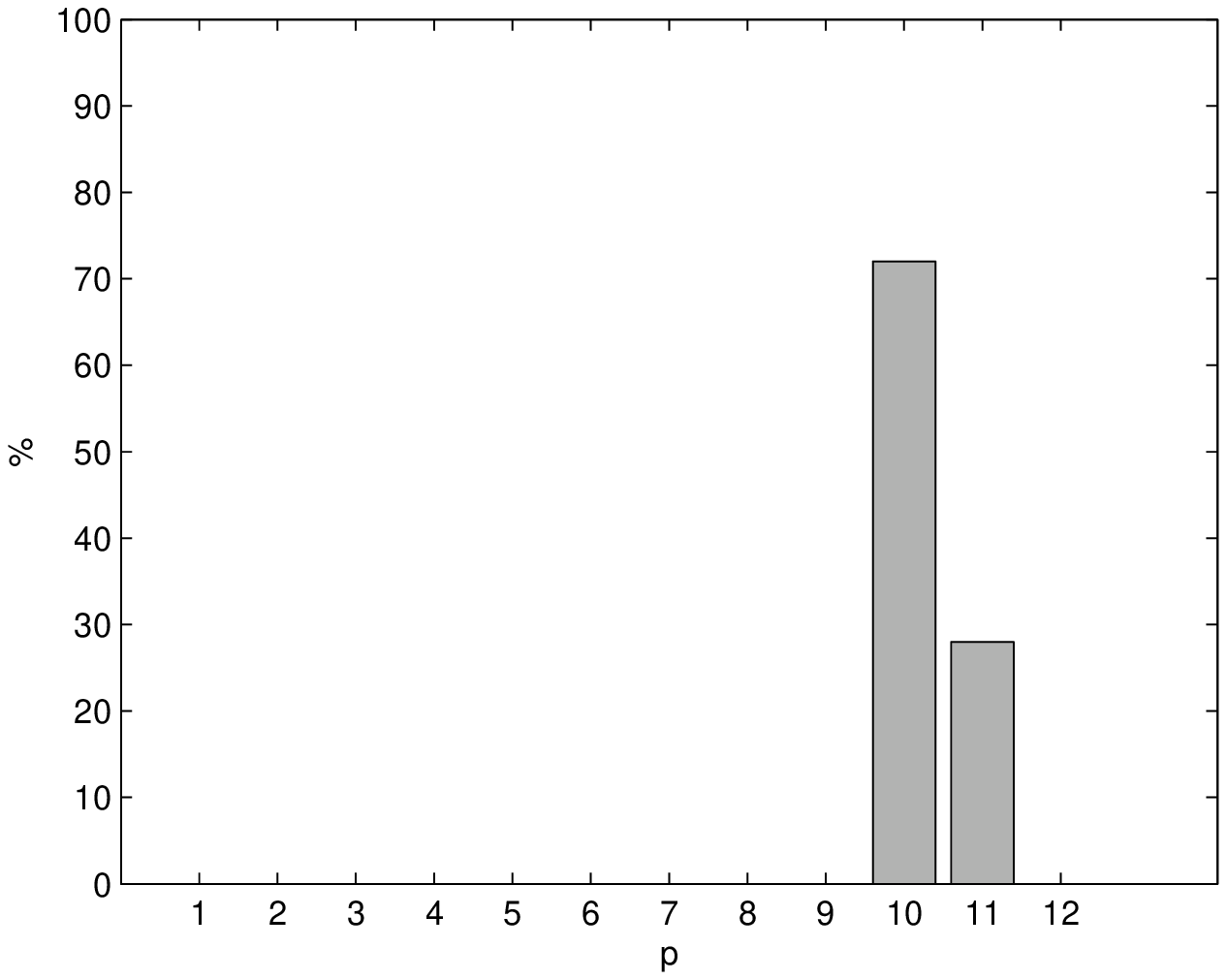}
\end{tabular}
\caption{Percentage of selecting respectively $(L,p)$ and $p$ by the BIC criterion for the proposed approach (left) and the regression mixture approach (right), with $K=2$}\label{bic}
\end{figure}

Preliminarily, the triplet $(K,L,p)$ for the proposed approach is tuned using the BIC criterion as follows: (i) twenty-five sets of 50 time series are randomly generated with $\sigma^2_k=2$ ; (ii) the proposed algorithm is run on each data set, with $K \in\{1,\ldots,K_{max}\}$, $L \in\{ 1,\ldots,L_{max}\}$ and $p\in\{1,\ldots,p_{max}\}$ ; (iii) the selection rate for each triplet $(K,L,p)$ over the 25 random samples is computed as a percentage. The model with the highest percentage of selections is the one with $(K,L,p)=(2,3,3)$.  The same strategy was applied to the regression mixture approach, where the pair $(K,p)=(2,10)$ was found to have the highest percentage of selections. Figure \ref{bic} shows the percentages obtained with the two algorithms, only for $K=2$.

Using the optimal numbers of clusters, segments and polynomial orders computed above, the two algorithms are then compared. Table \ref{results_cas1} gives the obtained misclassification percentages and intra-cluster inertia averaged over 25 random samples. The overall performance of the proposed algorithm is seen to be better than that of the regression mixture EM algorithm.
\begin{table}[htbp]
\centering
\caption{Misclassification rate and intra-cluster inertia obtained with the two compared algorithms}\label{results_cas1}
\begin{tabular}{ccc}
\hline &Misclassification percentage& Intra-cluster inertia\\\hline
Proposed approach &0 &$1.20 \times 10^4$ \\\hline
Regression mixture & 0.08&$2.25 \times 10^4$\\\hline
\end{tabular}
\end{table}

\begin{figure}[htbp]
\centering
\begin{tabular}{ccc}
\includegraphics[width=5.5cm]{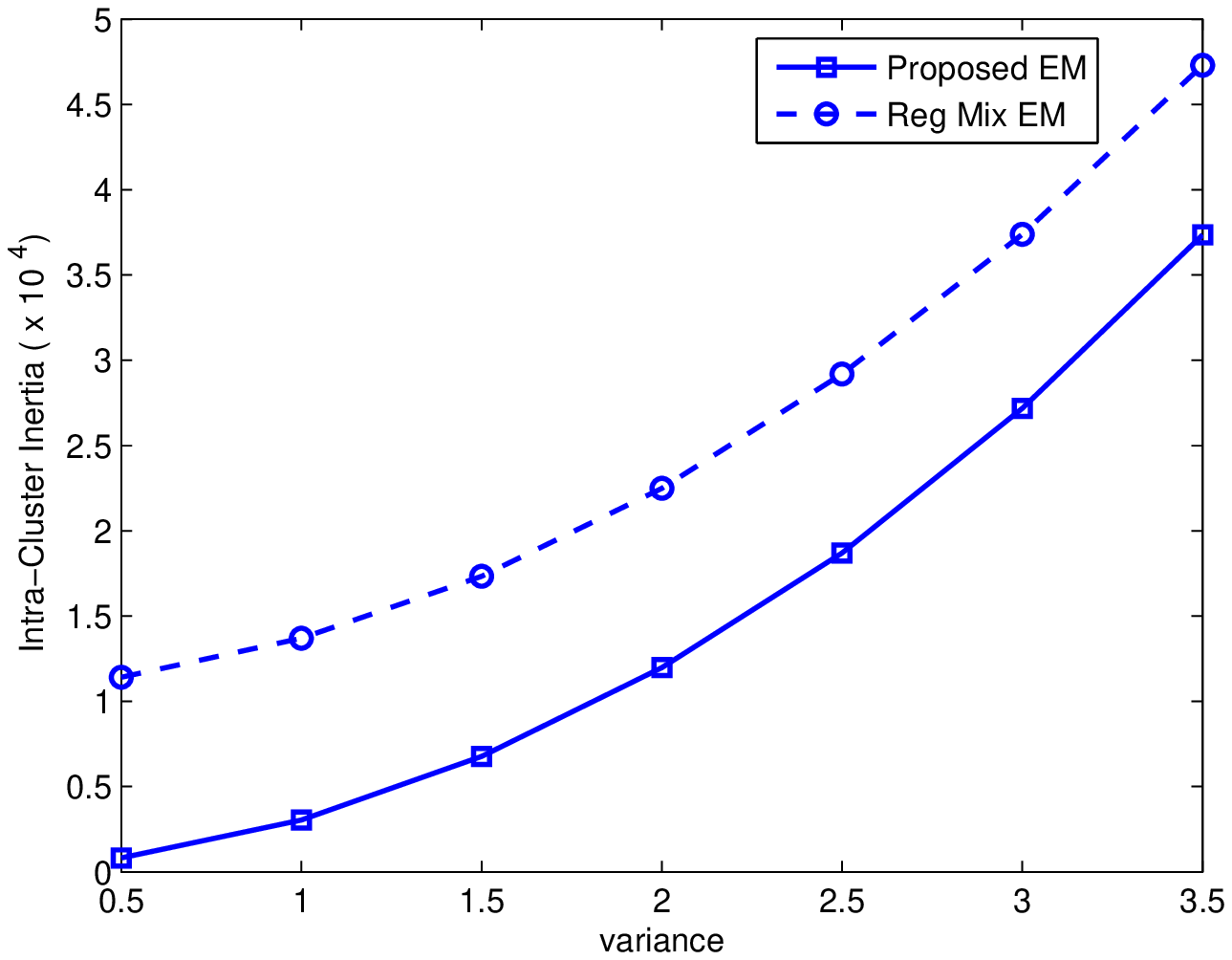}&
\includegraphics[width=5.5cm]{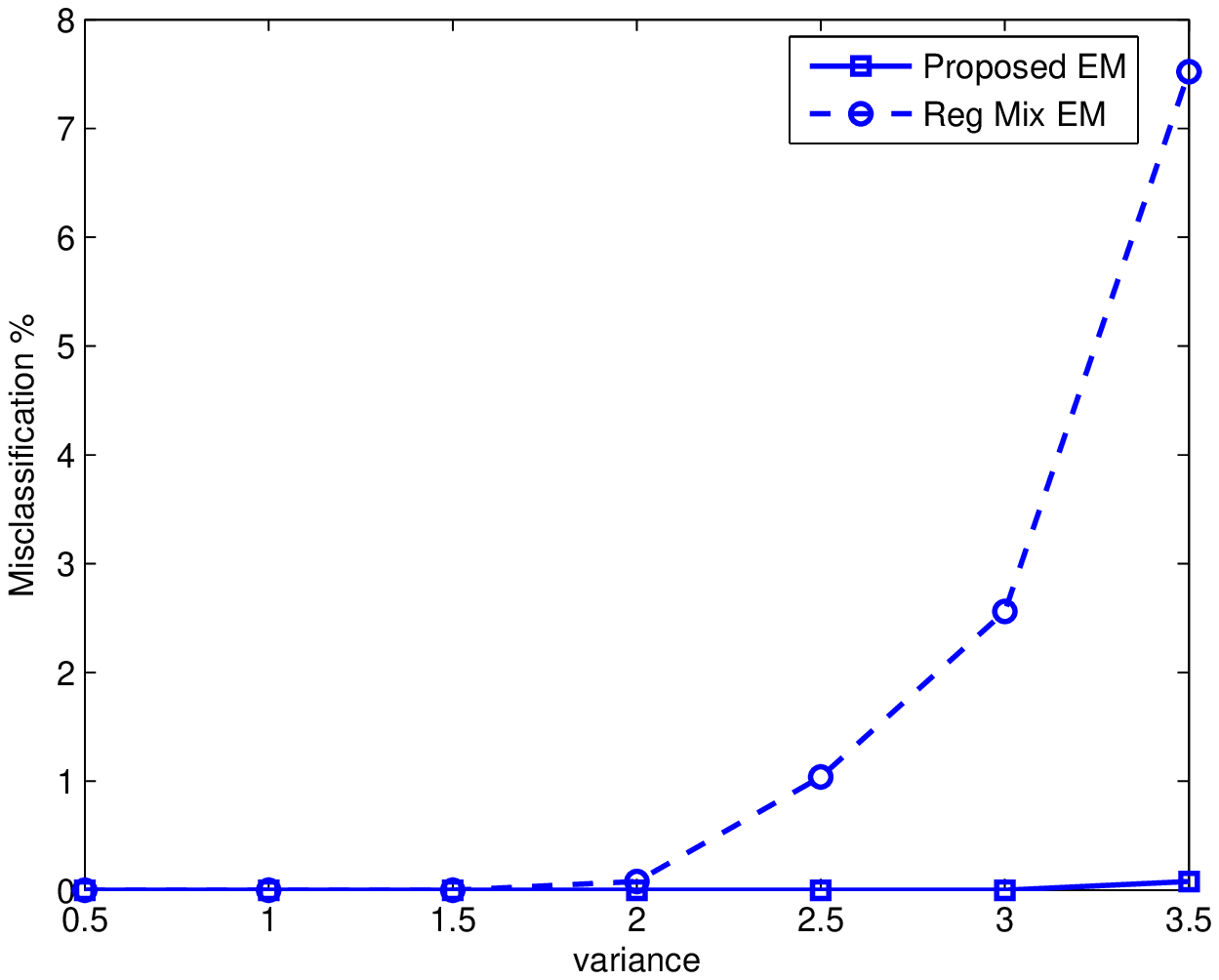}
\end{tabular}
\caption{Misclassification rate (left) and intra-cluster inertia (right) in relation to the noise variance, obtained with the proposed EM algorithm and the standard regression mixture EM algorithm}\label{results_cas12}
\end{figure}

\begin{figure}[htbp]
\centering
\begin{tabular}{cc}
Cluster 1& Cluster 2\\
\includegraphics[width=5cm,height=3cm]{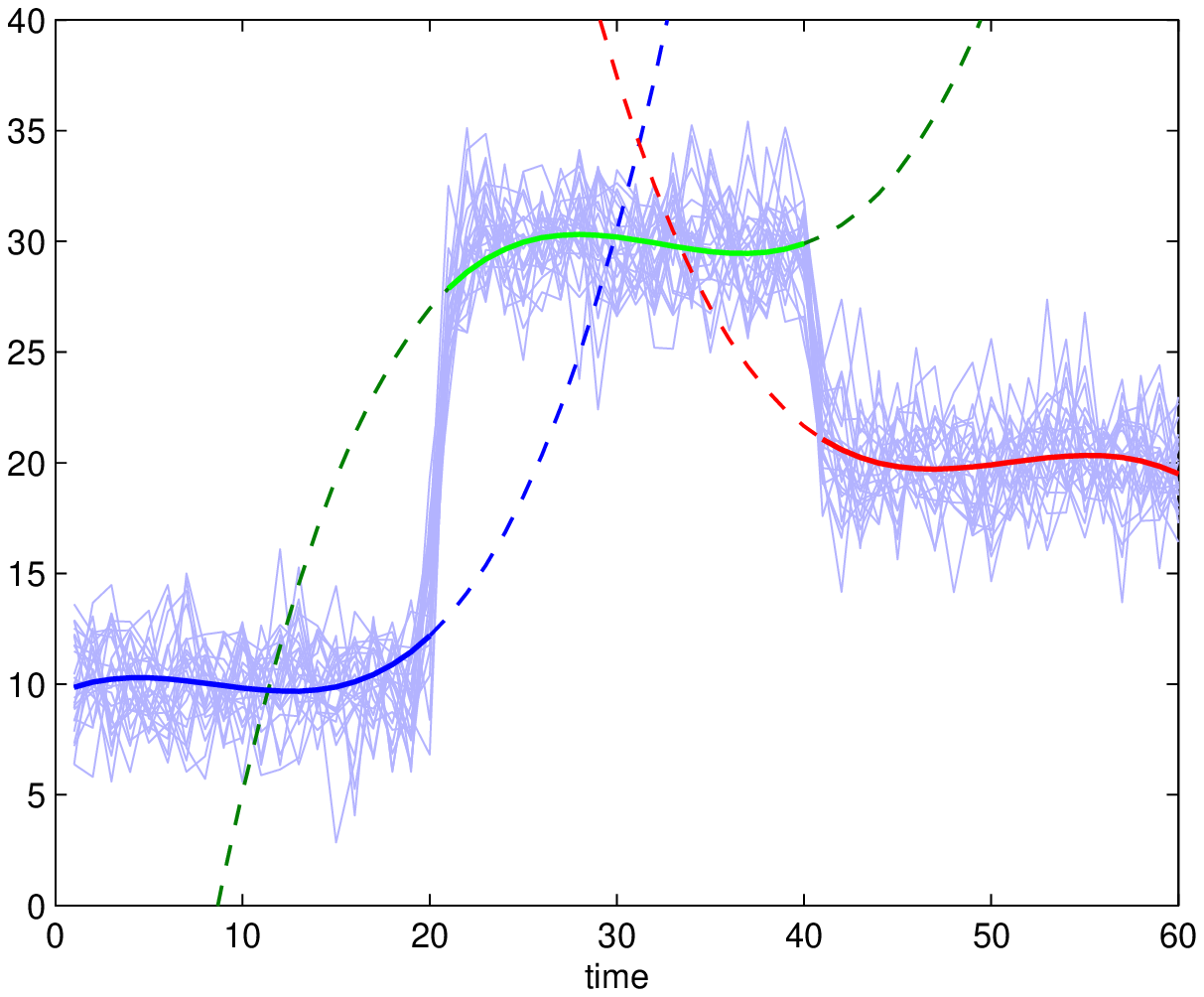} & \includegraphics[width=5cm,height=3cm]{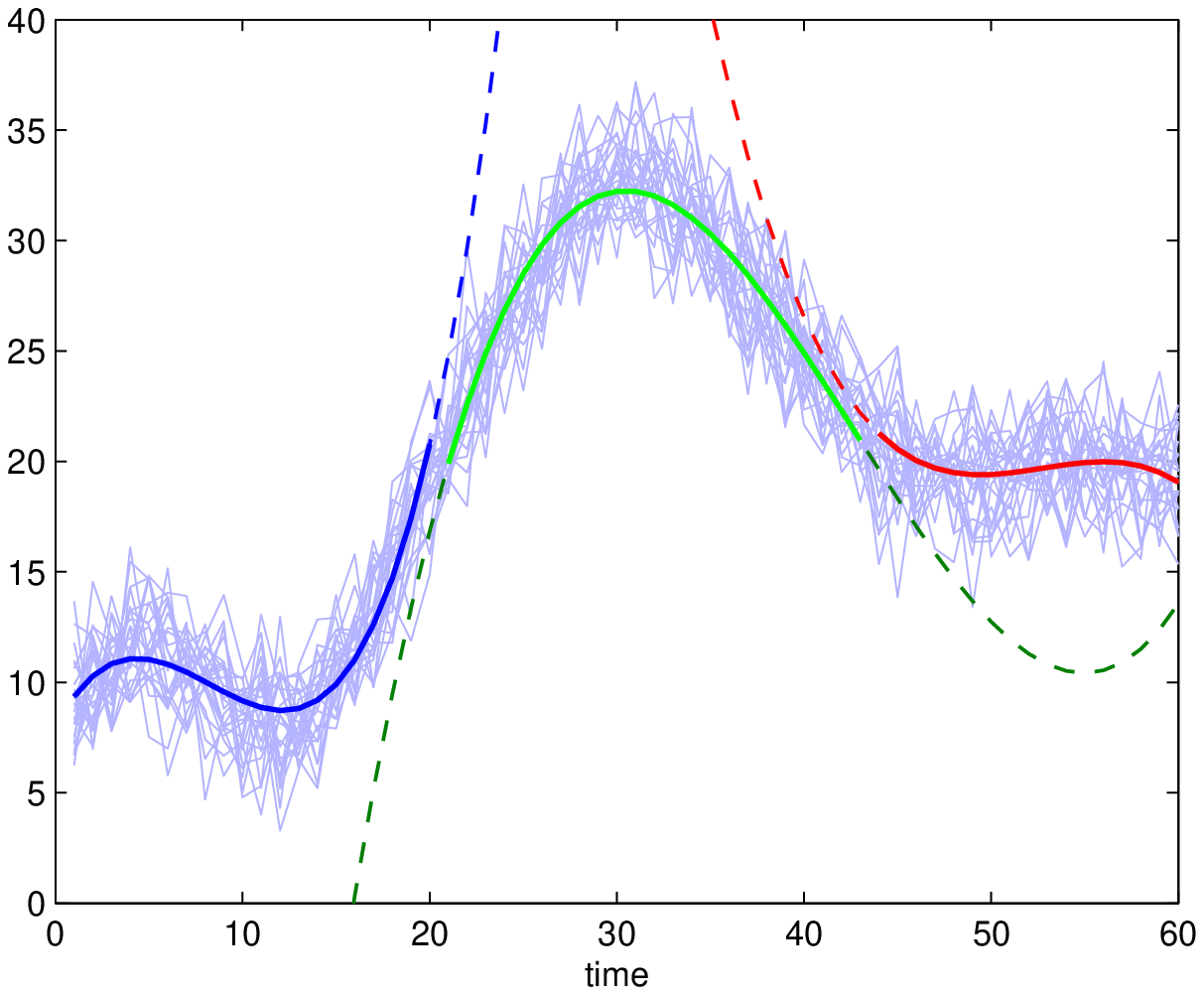}\\
\includegraphics[width=5cm,height=1.5cm]{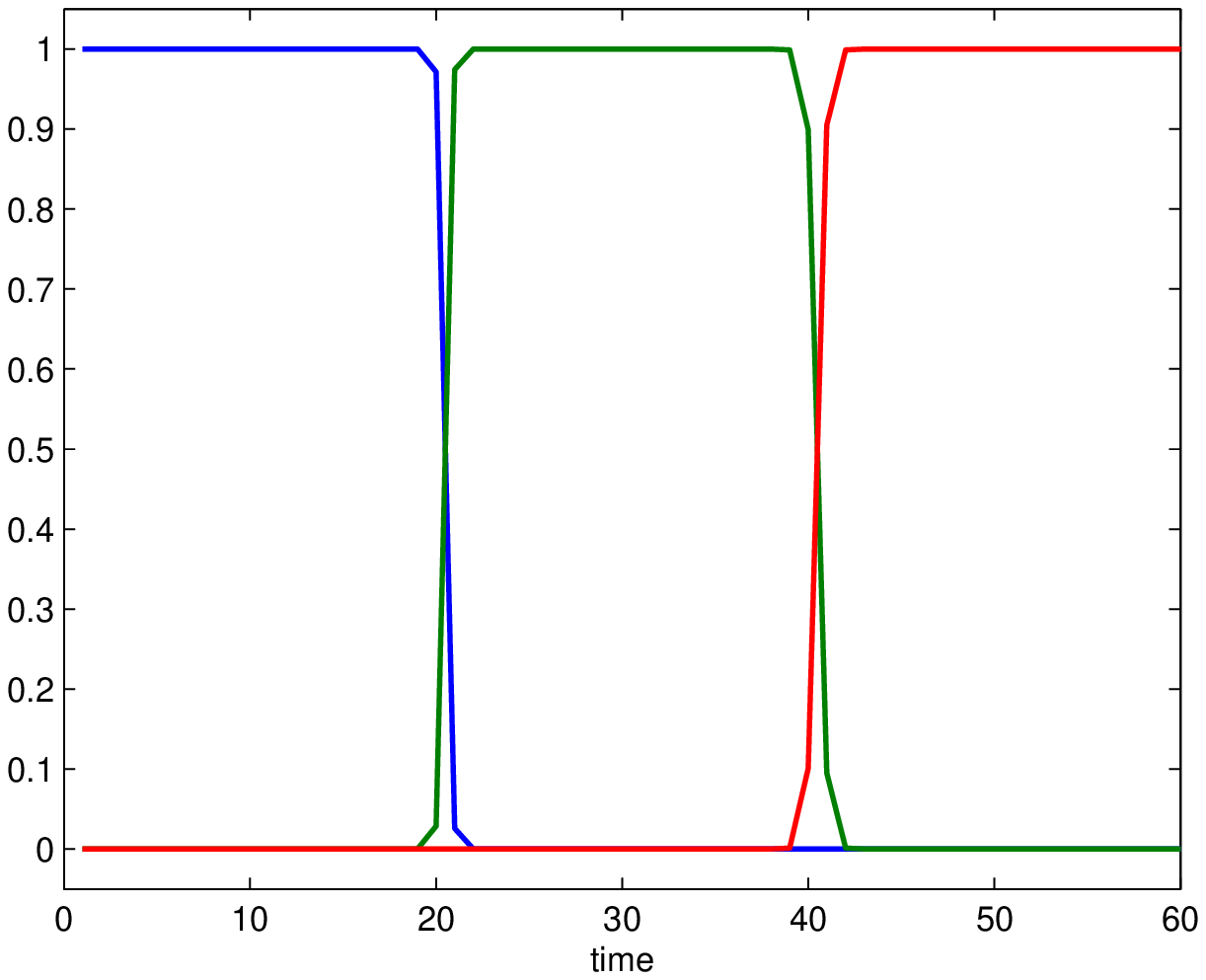}&
\includegraphics[width=5cm,height=1.5cm]{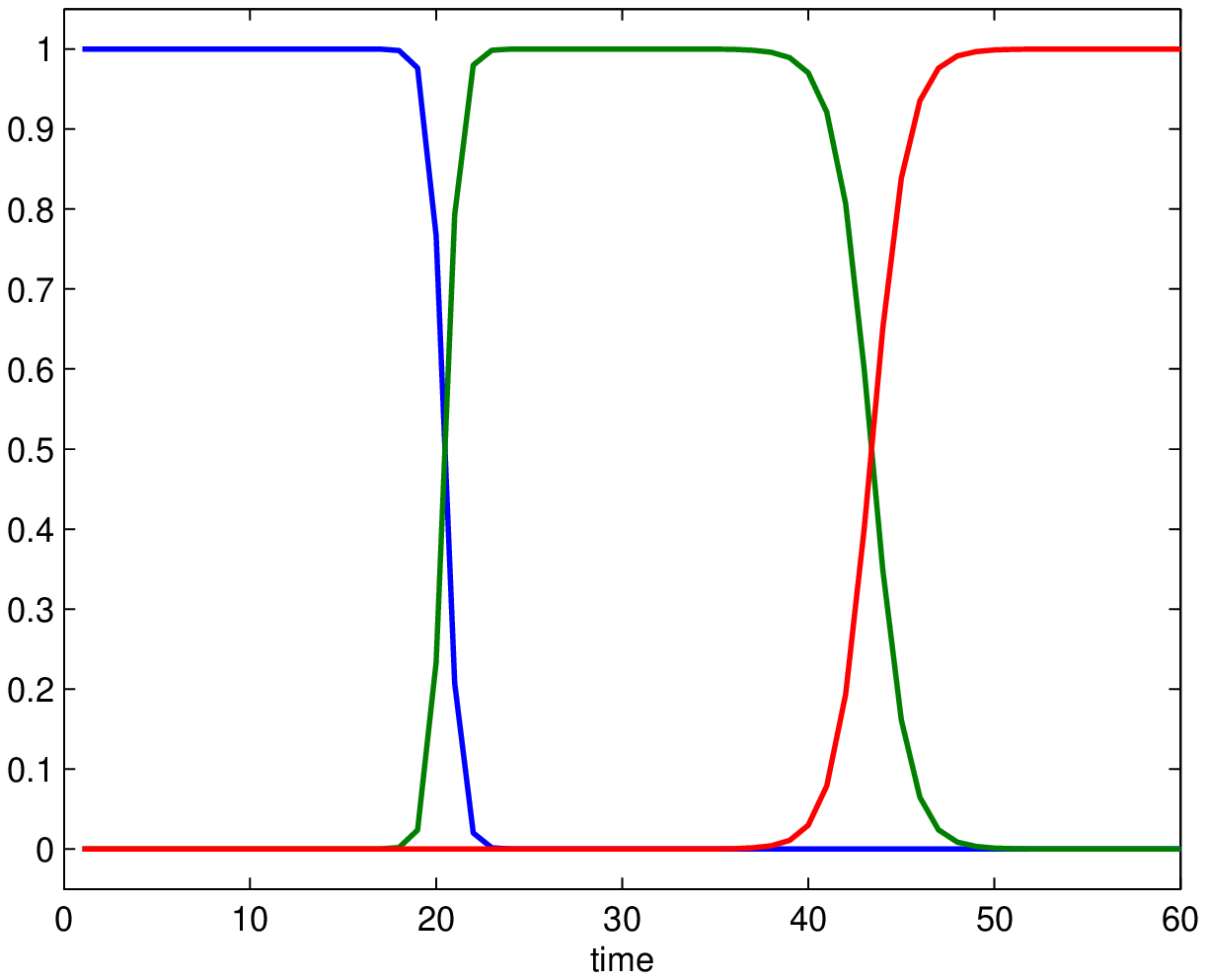}\\
\includegraphics[width=5cm,height=3cm]{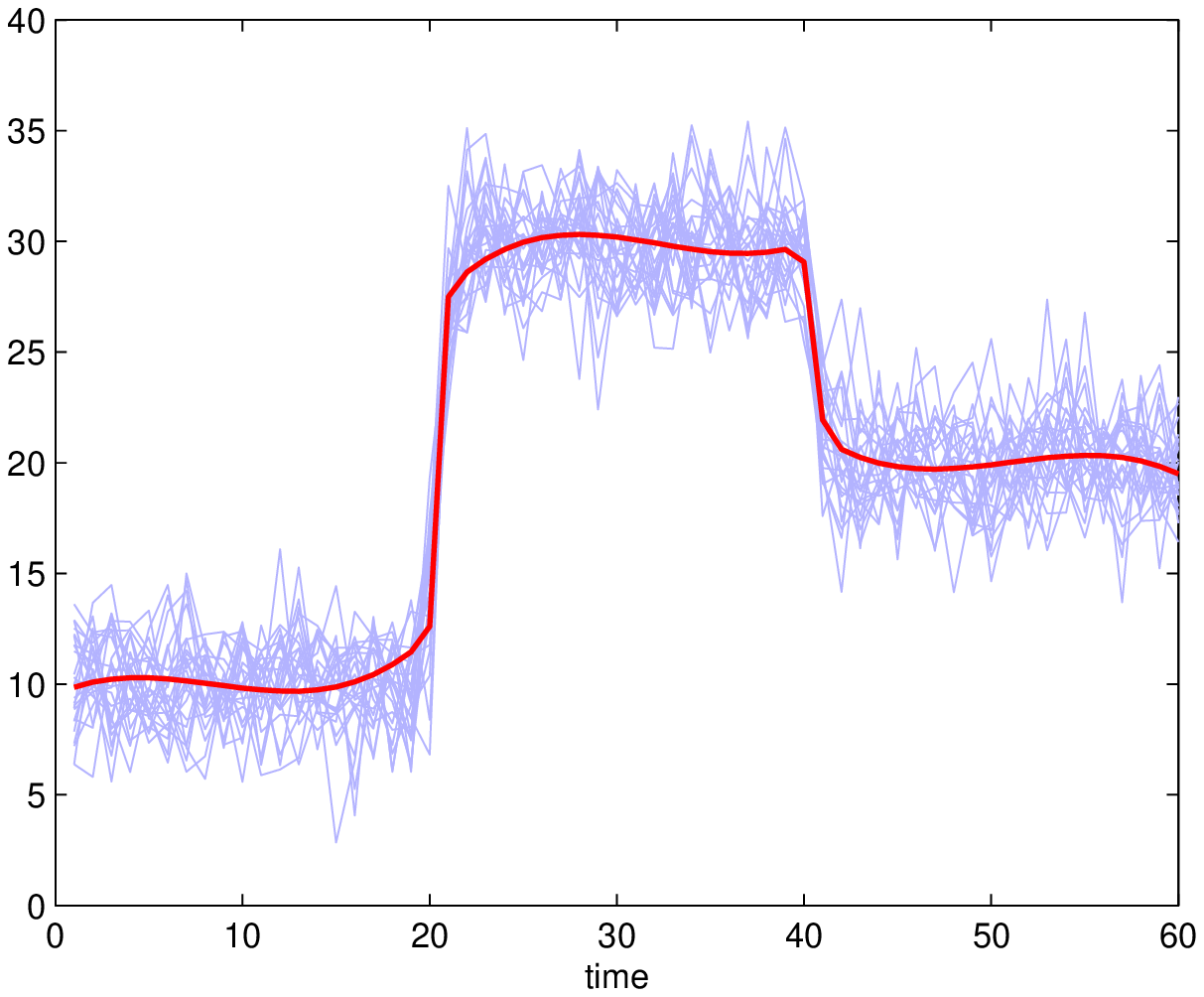}&
\includegraphics[width=5cm,height=3cm]{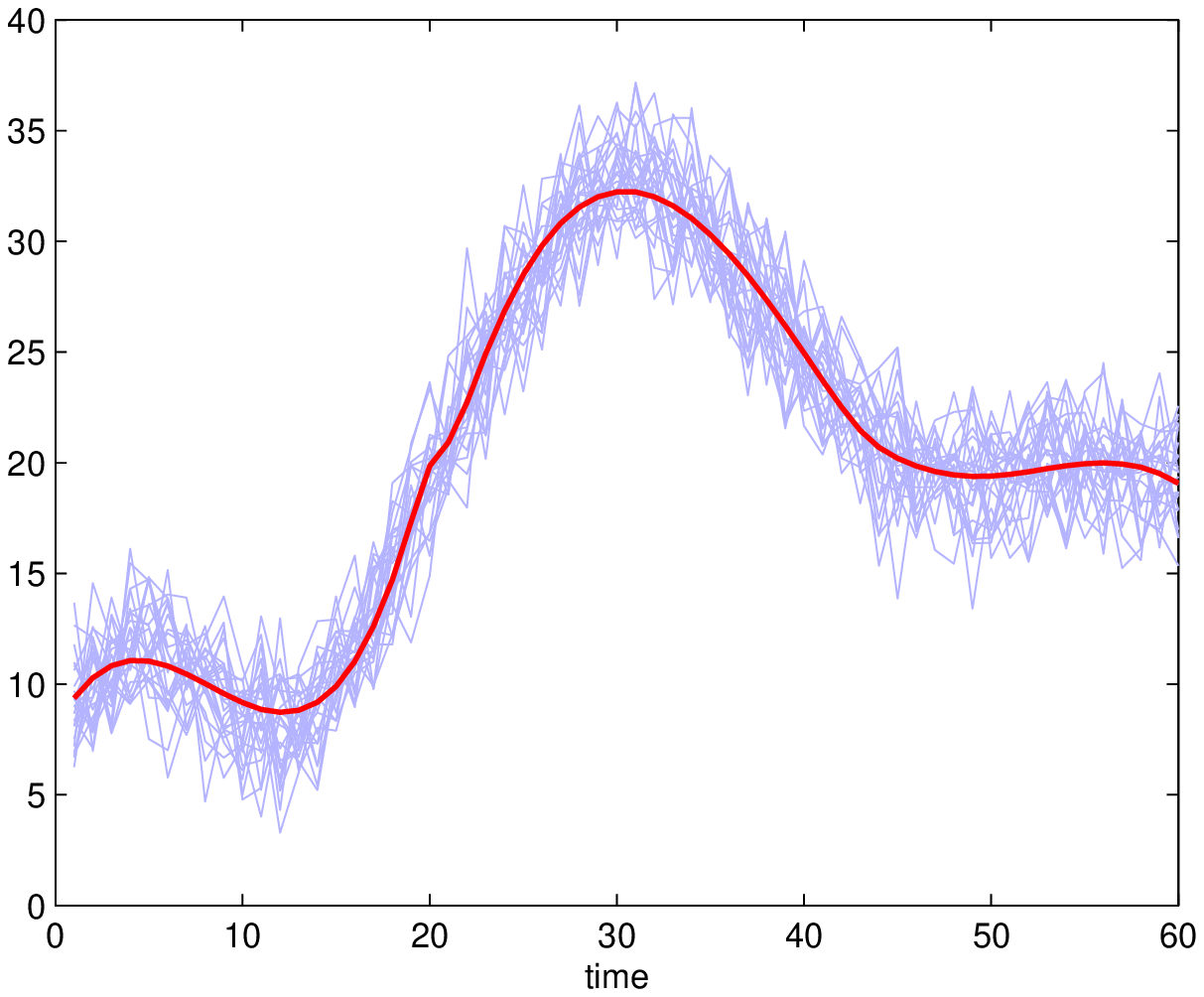}
\end{tabular}
\caption{Clustering results provided by the proposed EM algorithm applied with $K=2$, $L=3$ and $p=3$: clusters with their estimated polynomials (top), logistic probabilities (middle), clusters with their mean series (bottom)} \label{affichage_clust_our}
\end{figure}

\begin{figure}[htbp]
\centering
\begin{tabular}{cc}
Cluster 1 & Cluster 2\\
\includegraphics[width=5cm,height=3cm]{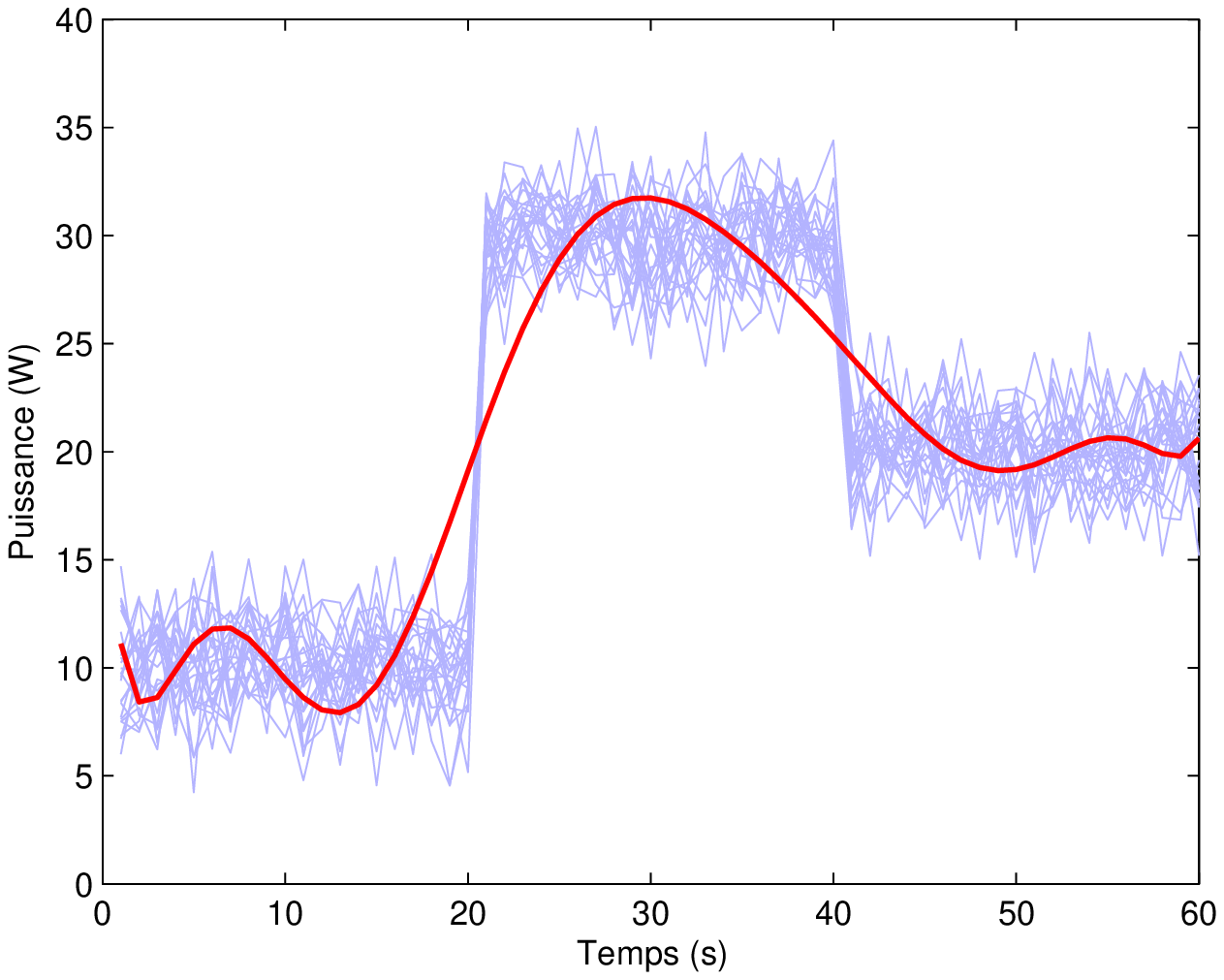} & \includegraphics[width=5cm,height=3cm]{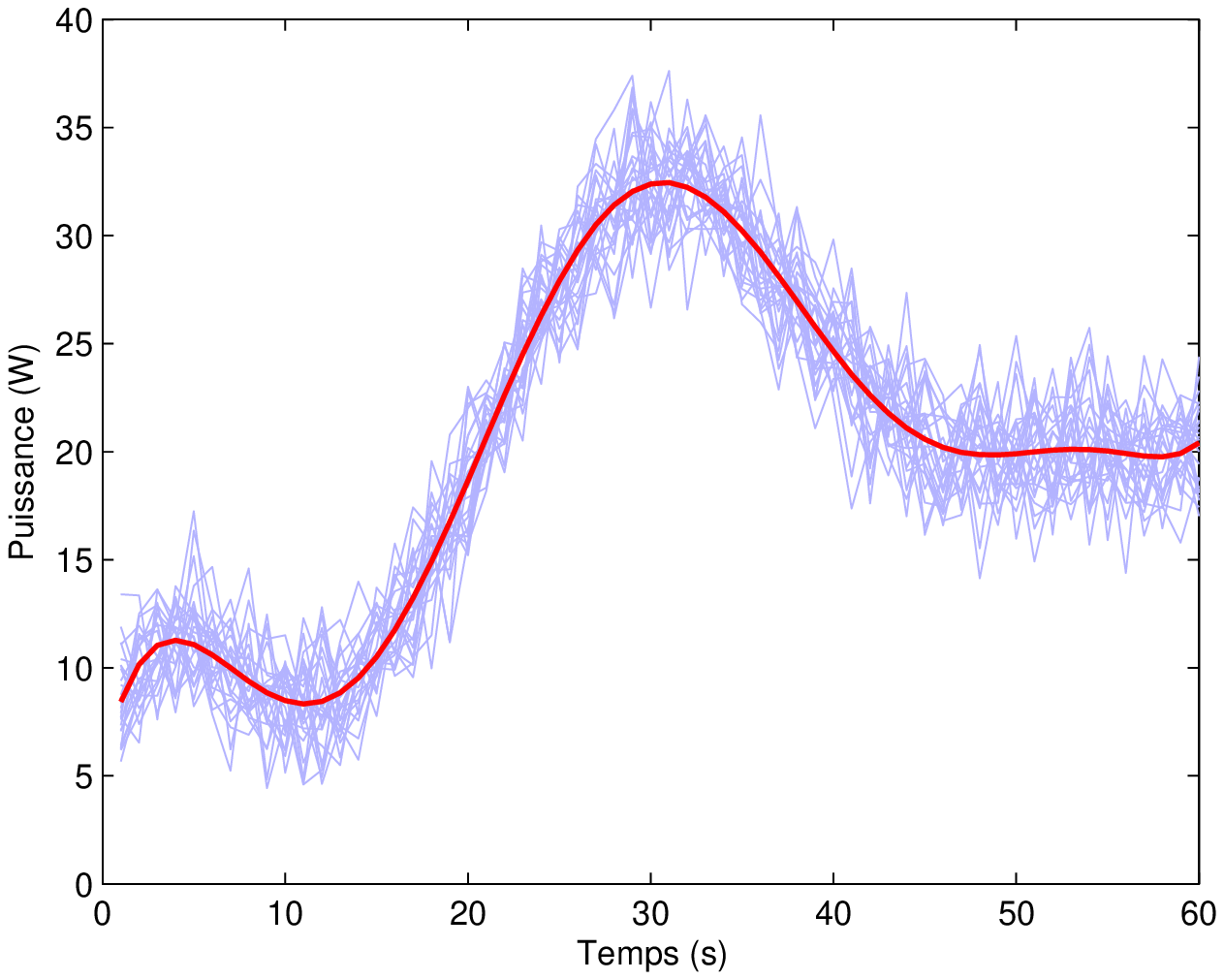}
\end{tabular}
\caption{Clusters and mean series estimated by the regression mixture EM algorithm applied with $(K,p)=(2,10)$}\label{affichage_clust_regmix}
\end{figure}

Figure \ref{results_cas12} shows the misclassification percentage and the intra-cluster inertia (averaged over 25 different random samples of time series) in relation to the variance $\sigma^2_k$, obtained with the proposed algorithm and the regression mixture EM algorithm. The proposed algorithm is seen to outperform its competitor. Although the misclassification percentages of the two approaches are close in particular for $\sigma^2_k\leq2$, the intra-cluster inertia differs from about $10^4$.  Misclassification provided by the regression mixture EM algorithm increases for variances greater than 2.5. The intra-cluster inertia obtained by the two approaches naturally increases with the variance level, but the proposed approach performs better than its competitor. Examples of clustering results provided by the proposed approach are displayed in figure \ref{affichage_clust_our}. It will be observed that our approach is also capable of modeling the cluster 2, whose mean series is a polynomial of degree 8, by means of three polynomials of order 3 weighted by logistic functions. Figure \ref{affichage_clust_regmix} illustrates that the regression mixture model, in contrast to the proposed model, cannot accurately model cluster 1, whose series are subject to changes in regime.

\subsection{Experiments using real world data}
\begin{figure}[htbp]
\centering
\includegraphics[width=7cm]{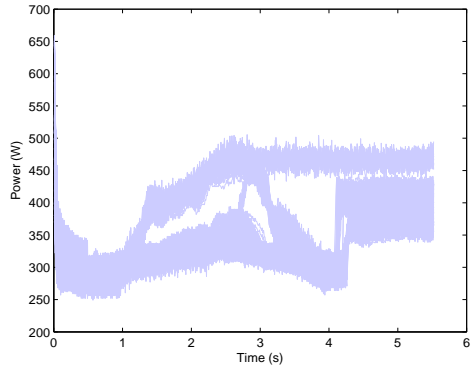}
\caption{Electrical power consumption time series acquires during $n=140$ switch operations} \label{real signals}
\end{figure}

As mentioned in the introduction, the main motivation behind this study was diagnosing problems in
the rail switches that allow trains to change tracks at junctions. An important preliminary task
in the diagnostic process is the automatic identification of groups of switching
operations that have similar characteristics, by analyzing time series of electrical power consumption acquired during switching operations. The specificity of the time series to be analyzed in this context is that they are subject to various changes in regime as a result of the mechanical movements involved in a switching operation. We accomplished this clustering task using our EM algorithm, designed for estimating the parameters of a mixture of hidden process regression models. We compared the proposed EM algorithm to the regression mixture EM algorithm previously described, on a data set of $n=140$ time series (see figure \ref{real signals}). This data set is composed of four clusters identified by an expert:
a defect-free cluster (35 time series),
a cluster with a minor defect (40 time series),
a cluster with a type 1 critical defect (45 time series) and
a cluster with a type 2 critical defect (20 time series).

\begin{figure}[htbp]
\centering
\begin{tabular}{cccc}
Cluster 1 & Cluster 2 & Cluster 3 & Cluster 4 \\
\includegraphics[width=2.6cm]{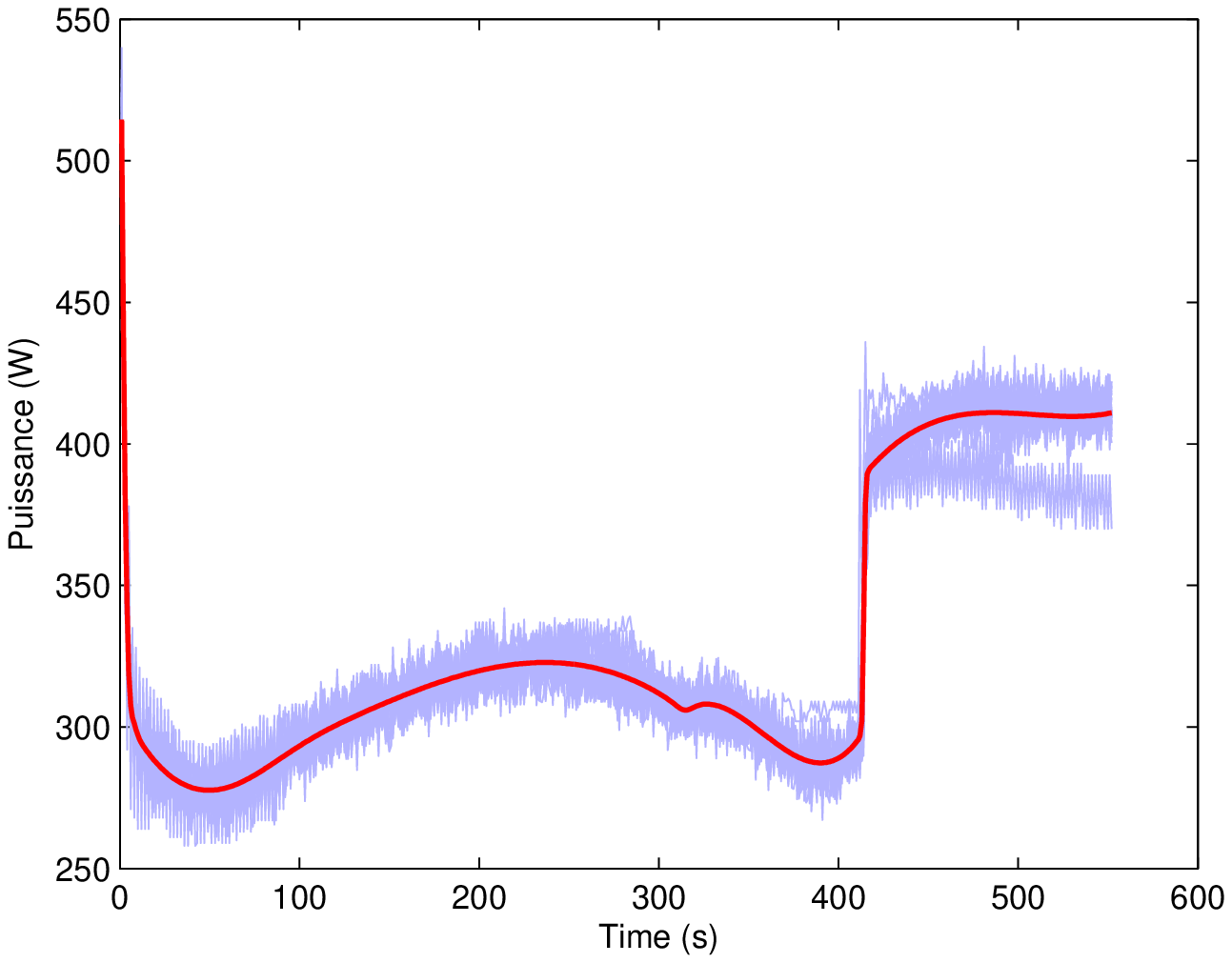}&\includegraphics[width=2.6cm]{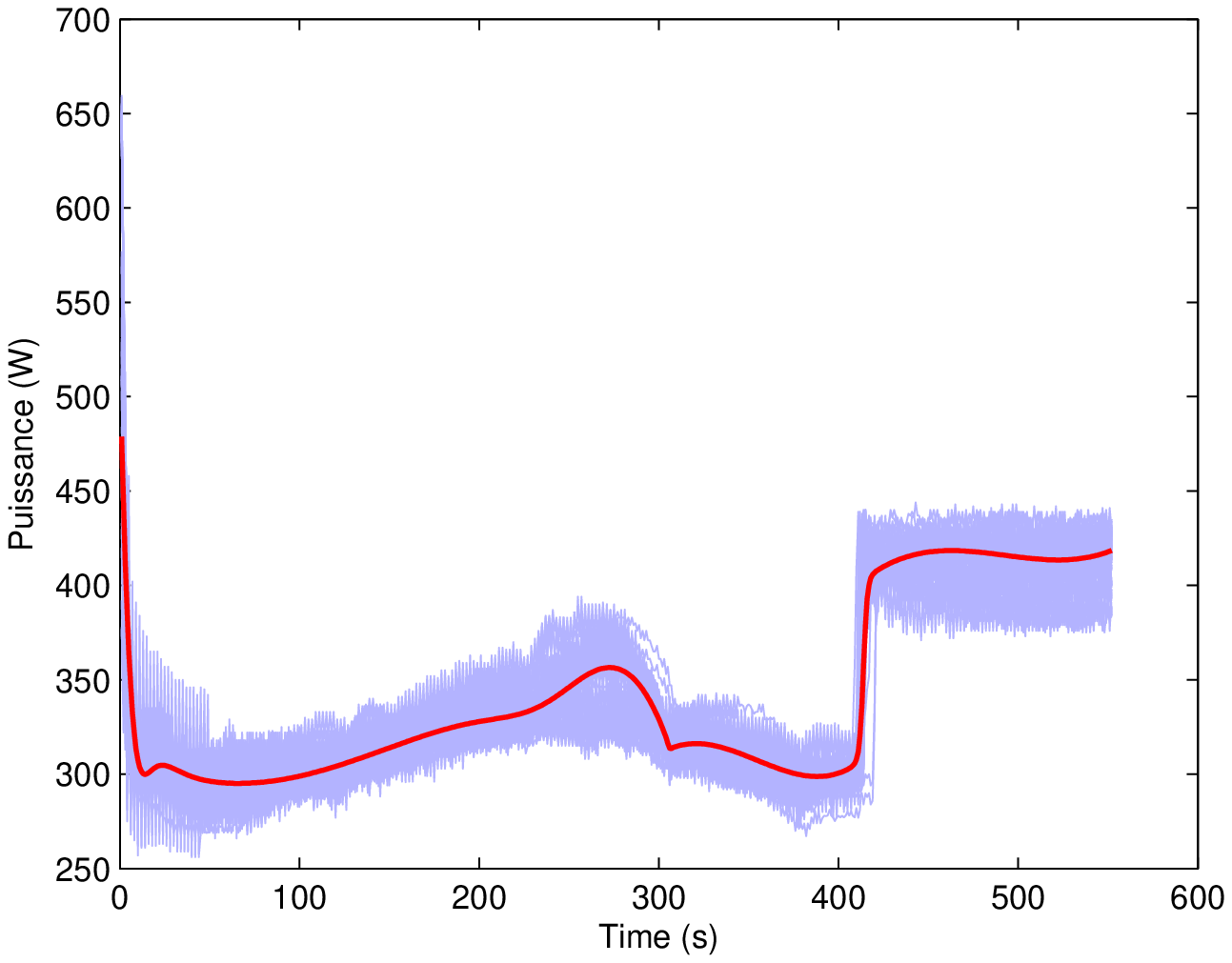}&\includegraphics[width=2.6cm]{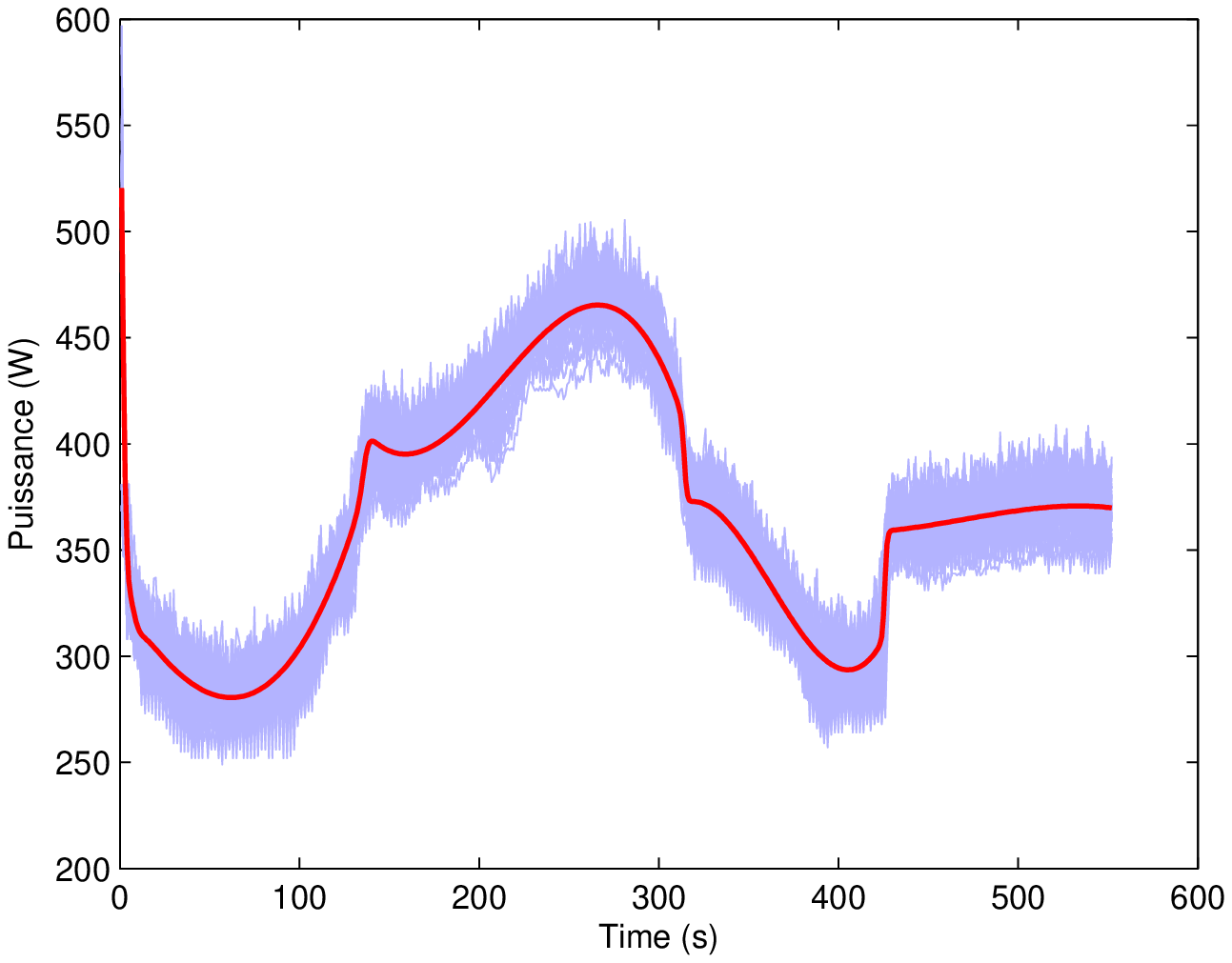}&\includegraphics[width=2.6cm]{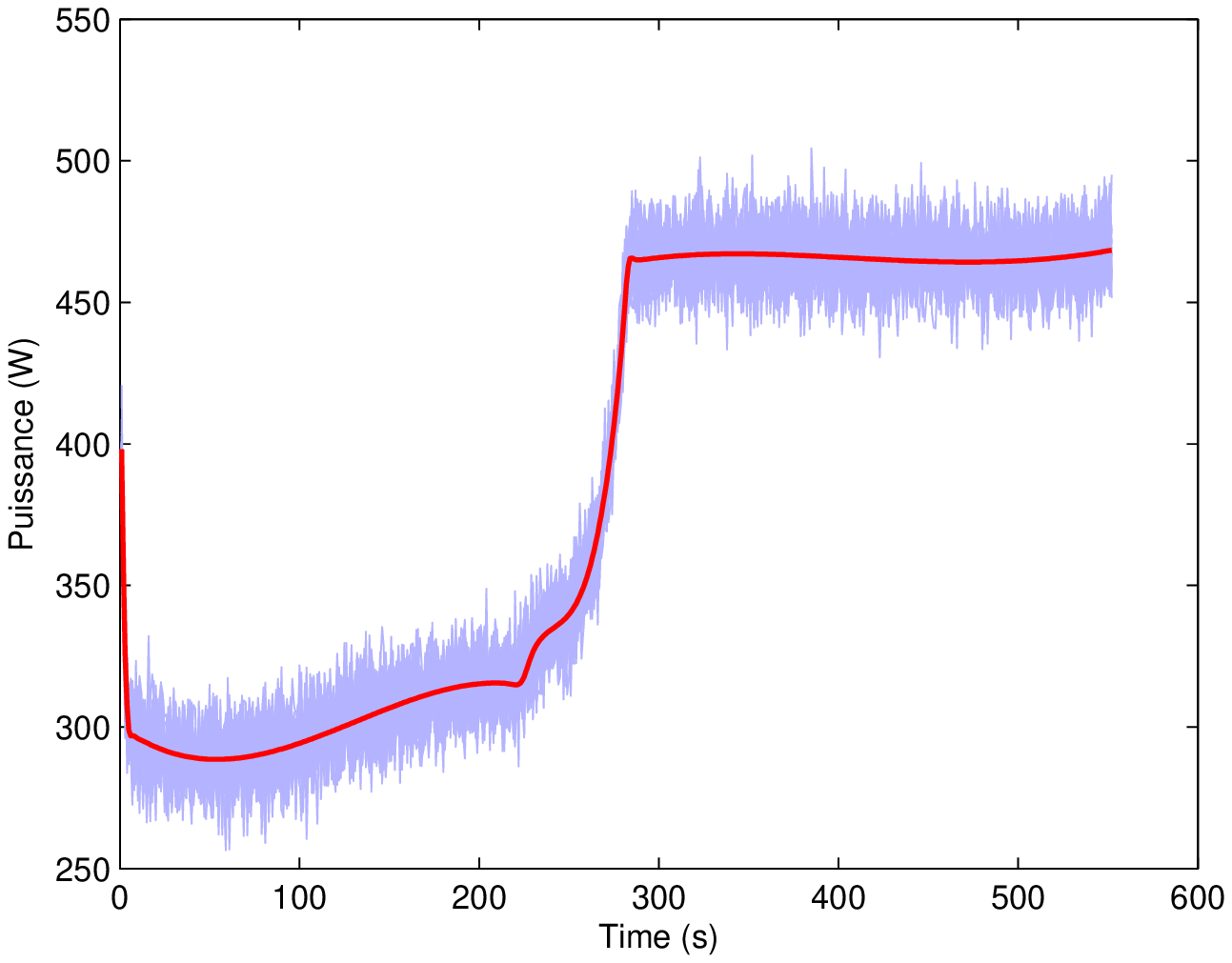}\\
\includegraphics[width=2.6cm]{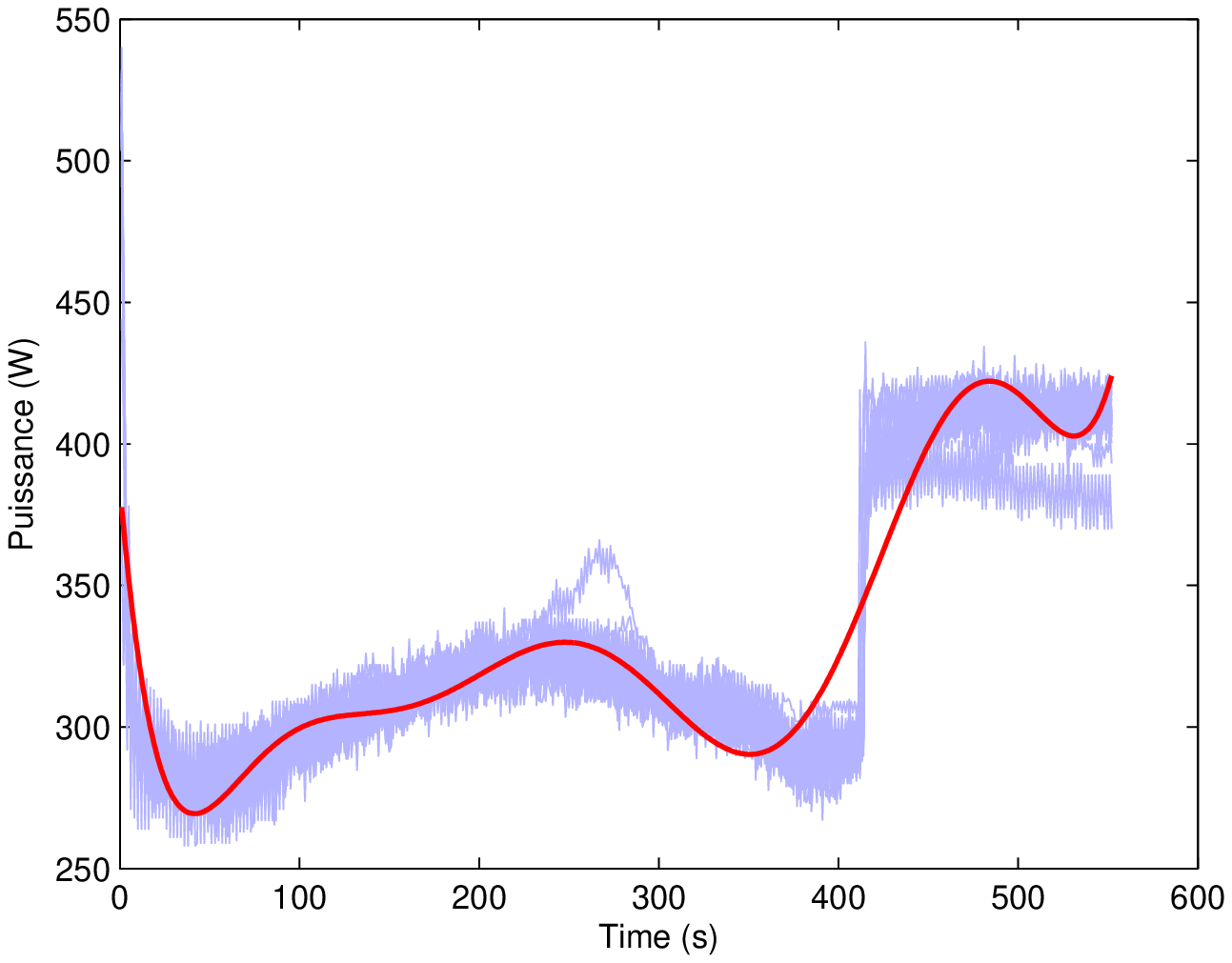}&\includegraphics[width=2.6cm]{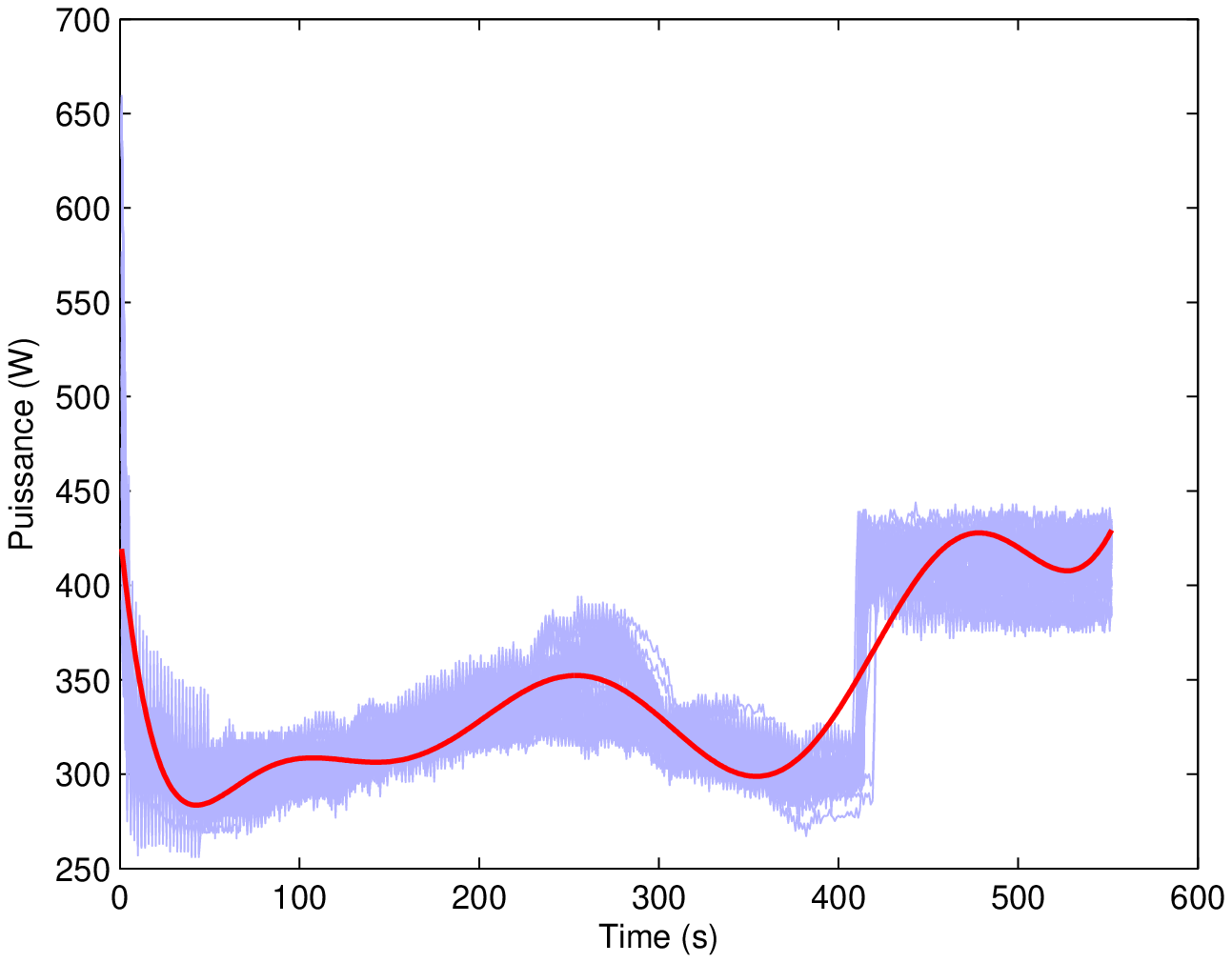}&\includegraphics[width=2.6cm]{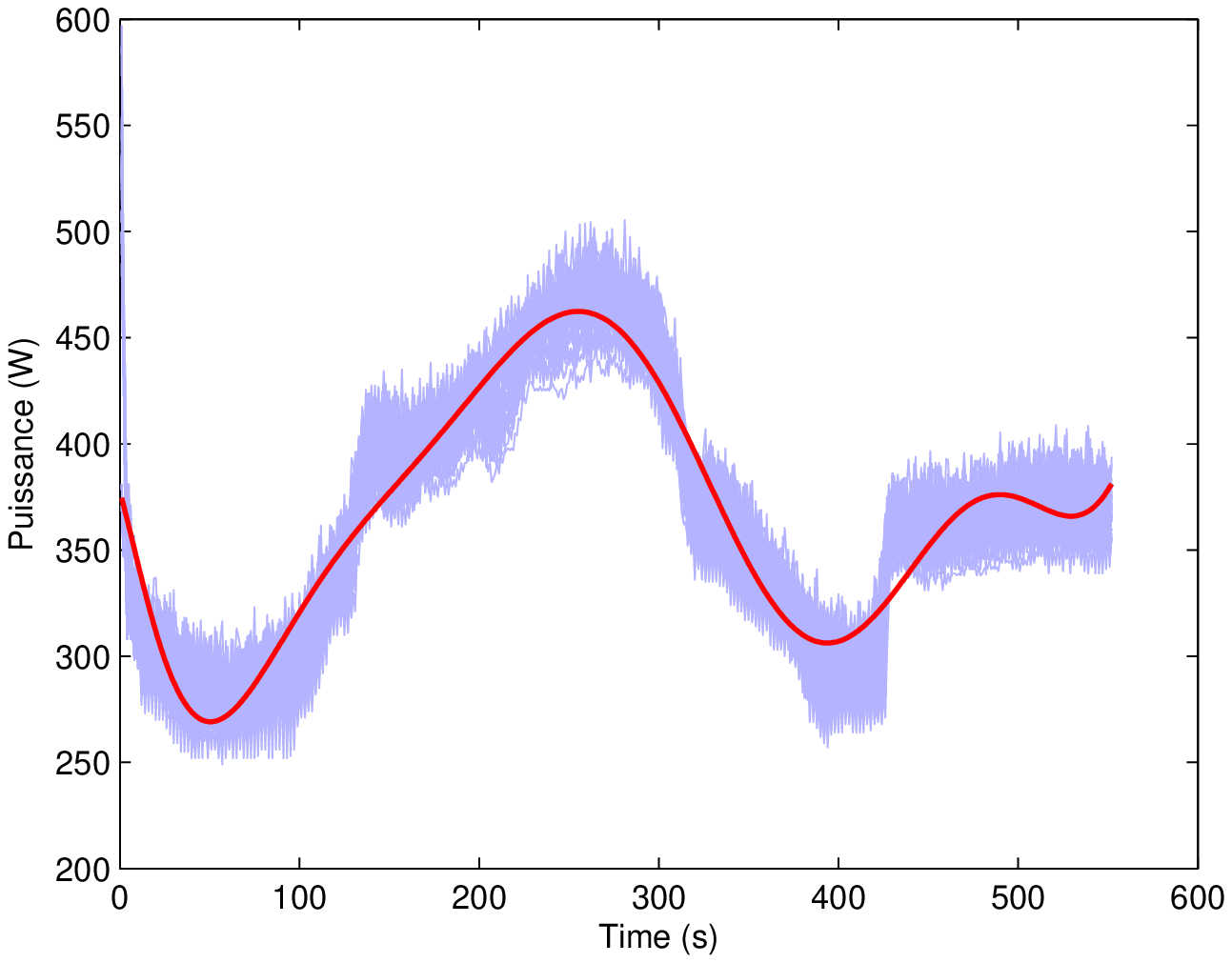}&\includegraphics[width=2.6cm]{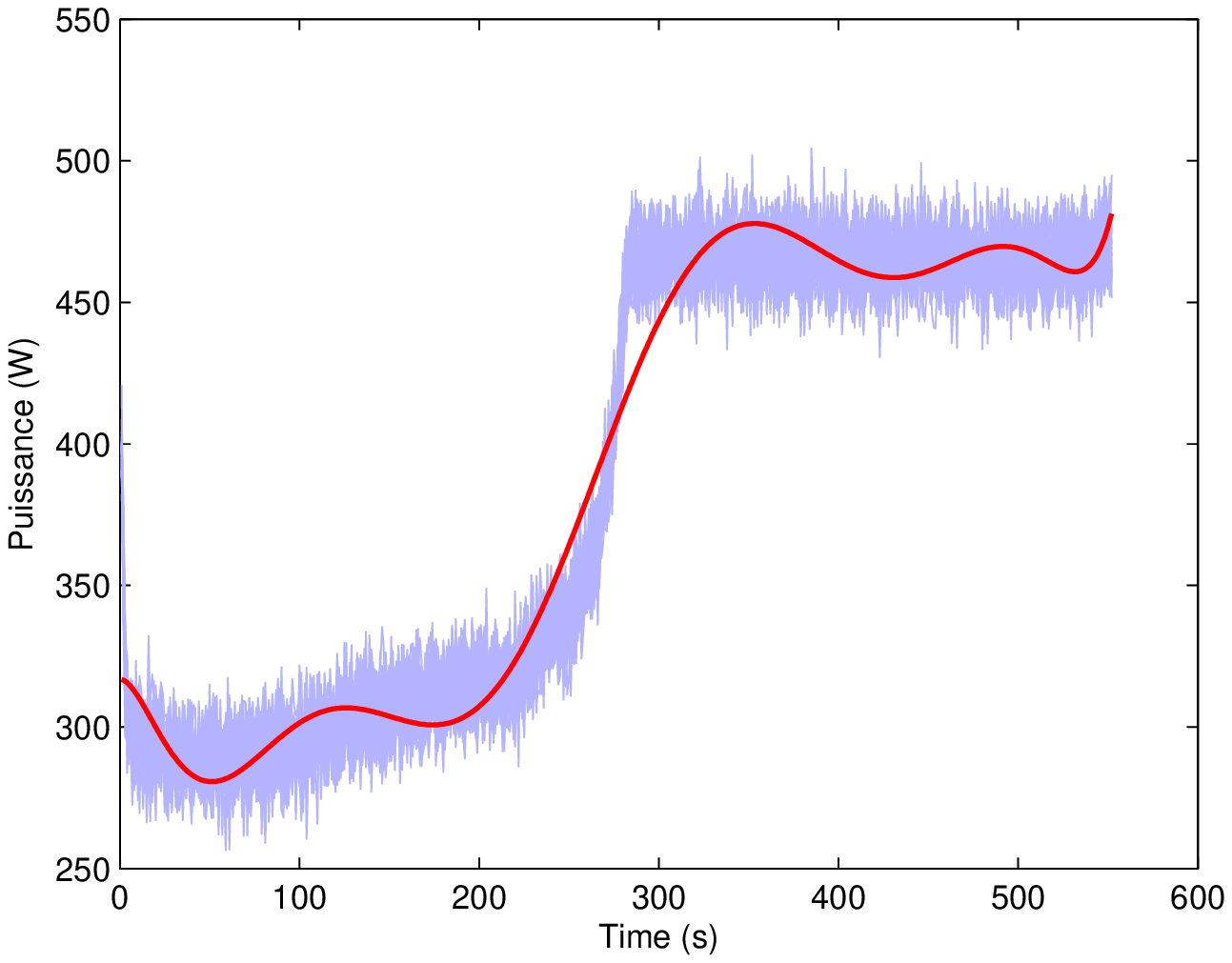}\\
\end{tabular}
\caption{Clusters and mean series estimated by the proposed EM algorithm (top) and the regression mixture EM algorithm (bottom)} \label{affichage_clust_reels}
\end{figure}

The number of regression components of the proposed algorithm was set to $L=5$ in accordance with the number of mechanical phases in a switching operation, and the degree of the polynomial regression $p$ was set to 3, which is more appropriate for the different regimes in the time series. The polynomial order for the regression mixture approach was set to $p=10$ which, in practice, gives the best error rates. For all the compared algorithms the number of clusters was set to $K=4$. Table \ref{results_reels} shows the misclassification error rates and the corresponding intra-cluster inertia. It can be seen that the proposed regression approach provides the smallest intra-cluster error and misclassification rate. Figure \ref{affichage_clust_reels} displays the clusters provided by the three compared algorithms and their estimated mean series.

\begin{table}[htbp]
\centering
\caption{Error obtained for the three compared approaches}\label{results_reels}
\begin{tabular}{ccc}
\hline
& Regression mixture EM & Proposed EM\\\hline
Misclassification \% &11.42&9.28 \\ \hline Intra-cluster inertia &$2.6583\times10^7$&$1.1566\times10^7$\\\hline
\end{tabular}
\end{table}

\section{Conclusion}
A new mixture model-based approach for the clustering of univariate time series with changes in regime has been
proposed in this paper. This approach involves modeling each cluster using a particular regression model whose polynomial coefficients vary over time according to a discrete hidden process. The transition between regimes is smoothly controlled by logistic functions. The model parameters are
estimated by the maximum likelihood method, solved by a dedicated
Expectation-Maximization (EM) algorithm. The proposed approach can also
be regarded as a clustering approach which operates by finding groups of
time series having common changes in regime. The Bayesian Information Criterion (BIC) is used to determine the numbers of clusters and segments, as well as the regression order. The experimental results, both from simulated time series and from a real-world application, show that the proposed approach is an efficient means for clustering univariate time series with changes in regime.

\acknowledgement{The authors wish to thank M. Marc Antoni of SNCF for the data he provided and for the support he has given them.}

\appendix
\section{Convexity of the set $E_{k\ell}$}\label{appendix1}
The set $E_{k\ell}$ defined by:
\begin{equation*}
E_{k\ell} = \Big \{t\in[t_1;t_m]\ /\   \pi_{k\ell}(t;\bal_k)=\max_{1\leq h \leq L}\pi_{kh}(t;\bal_k) \Big\}.
\end{equation*}
is a convex set of $\mathds{R}$. In fact, we have the following equalities:
\begin{eqnarray*}
E_{k\ell} &=& \Big \{t\in [t_1;t_m]\ /\   \pi_{k\ell}(t;\bal_k)=\max_{1\leq h \leq L}\pi_{kh}(t;\bal_k) \Big\}
\\&=&  \Big\{t\in [t_1;t_m]\ /\  \pi_{kh}(t;\bal_k) \leq \pi_{k\ell}(t;\bal_k)  \ \  \mbox{for } h=1,\ldots,L\Big\}\\\\
&=&  \bigcap_{1\leq h \leq L}\ \Big\{t\in [t_1;t_m]\ /\  \pi_{kh}(t;\bal_k) \leq \pi_{k\ell}(t;\bal_k)  \Big\}\\
&=&\bigcap_{1\leq h \leq L}\ \Big \{t\in [t_1;t_m]\ /\  \ln \frac{\pi_{kh}(t;\bal_k)}{\pi_{k\ell}(t;\bal_k)} \leq 0  \Big \}
\end{eqnarray*}
From the definition of $\pi_{k\ell}(t;\bal_k)$ (see equation \ref{logit_function}), it can be easily verified that $\ln \frac{\pi_{kh}(t;\bal_k)}{\pi_{k\ell}(t;\bal_k)}$ is a linear function of $t$.
Consequently, $E_{k\ell}$ is convex, as the intersection of convexes parts of $\mathds{R}$.

\end{document}